\documentclass[11pt]{article}

\usepackage[utf8]{inputenc}
\usepackage[a4paper,margin=1.2in]{geometry}
\usepackage{graphicx}% Include figure files
\usepackage{dcolumn}% Align table columns on decimal point
\usepackage{bm}% bold math
\usepackage{array}
\usepackage{multirow}
\usepackage{soul}
\usepackage{color}
\usepackage[binary-units=true]{siunitx}
\usepackage[flushleft]{threeparttable}
\usepackage{setspace}
\usepackage[style=nature,url=false]{biblatex}
\usepackage{authblk}
\usepackage{xr}
\usepackage[mathlines]{lineno}% Enable numbering of text and display math
\title{3D diffractive imaging of nanoparticle ensembles using an X-ray laser}
\author[1,2,3,$\dagger$,*]{Kartik Ayyer}
\author[1,3,4,$\dagger$]{P.\ Lourdu Xavier}
\author[5]{Johan Bielecki}
\author[6]{Zhou Shen}
\author[6]{Benedikt J.\ Daurer}
\author[4]{Amit K.\ Samanta}
\author[4]{Salah Awel}
\author[5]{Richard Bean}
\author[4]{Anton Barty}
\author[7]{Tomas Ekeberg}
\author[4]{Armando D.\ Estillore}
\author[5]{Klaus Giewekemeyer}
\author[8]{Mark S.\ Hunter}
\author[9]{Richard A.\ Kirian}
\author[5]{Henry Kirkwood}
\author[5]{Yoonhee Kim}
\author[5]{Jayanath Koliyadu}
\author[3,10]{Holger Lange}
\author[5]{Romain Letruin}
\author[3,4,11]{Jannik L{\"u}bke}
\author[12]{Andrew J.\ Morgan}
\author[4,11]{Nils Roth}
\author[5]{Tokushi Sato}
\author[5]{Marcin Sikorski}
\author[10]{Florian Schulz}
\author[9]{John C.~H.\ Spence}
\author[4,5]{Patrik Vagovic}
\author[1,2,3,11]{Tamme Wollweber}
\author[4,11]{Lena Worbs}
\author[4]{Oleksandr Yefanov}
\author[1,2]{Yulong Zhuang}
\author[7,13]{Filipe R.~N.~C.\ Maia}
\author[3,4,14]{Daniel A.\ Horke}
\author[3,4,11,15]{Jochen K{\"u}pper}
\author[6,16]{N.\ Duane Loh}
\author[5,17]{Adrian P.\ Mancuso}
\author[3,4,11]{Henry N.\ Chapman}

\affil[1]{Max Planck Institute for the Structure and Dynamics of Matter, 22761 Hamburg, Germany}
\affil[2]{Center for Free-Electron Laser Science, 22761 Hamburg, Germany}
\affil[3]{The Hamburg Center for Ultrafast Imaging, Universit{\"a}t Hamburg, 22761 Hamburg, Germany}
\affil[4]{Center for Free-Electron Laser Science, DESY, 22607 Hamburg, Germany}
\affil[5]{European XFEL, 22869 Schenefeld, Germany}
\affil[6]{Center for BioImaging Sciences, National University of Singapore, Singapore 117557}
\affil[7]{Dept. of Cell and Molecular Biology, Uppsala University, 75124 Uppsala, Sweden}
\affil[8]{Linac Coherent Light Source, SLAC National Accelerator Laboratory, Menlo Park, 94025, USA}
\affil[9]{Department of Physics, Arizona State University, Tempe, AZ 85287, USA}
\affil[10]{Institute of Physical Chemistry, Universit{\"a}t Hamburg, 20146 Hamburg, Germany}
\affil[11]{Department of Physics, Universit{\"a}t Hamburg, 22761 Hamburg, Germany}
\affil[12]{Univ. of Melbourne, Physics, Victoria, 3010, Australia}
\affil[13]{NERSC, Lawrence Berkeley National Laboratory, Berkeley, CA 94720, USA}
\affil[14]{Institute for Molecules and Materials, Radboud University, 6525 AJ Nijmegen, Netherlands}
\affil[15]{Department of Chemistry, Universit{\"a}t Hamburg, 20146 Hamburg, Germany}
\affil[16]{Department of Physics, National University of Singapore, Singapore 117551}
\affil[17]{Department of Chemistry and Physics, La Trobe Institute for Molecular Science, La Trobe University, Melbourne, Victoria 3086, Australia}

\affil[$\dagger$]{These authors contributed equally.}
\affil[*]{kartik.ayyer@mpsd.mpg.de}

\date{}

\begin{document}

\maketitle

\begin{abstract}
\noindent
We report the 3D structure determination of gold nanoparticles (AuNPs) by X-ray single particle imaging (SPI). Around 10 million diffraction patterns from gold nanoparticles were measured in less than 100 hours of beam time, more than 100 times the amount of data in any single prior SPI experiment, using the new capabilities of the European X-ray free electron laser which allow measurements of 1500 frames per second. A classification and structural sorting method was developed to disentangle the heterogeneity of the particles and to obtain a resolution of better than \SI{3}{\nano\meter}. With these new experimental and analytical developments, we have entered a new era for the SPI method and the path towards close-to-atomic resolution imaging of biomolecules is apparent.
\end{abstract}

The determination of the structures of biomolecules at atomic resolution requires bright sources of radiation, which are unfortunately also energetic enough to degrade the object under observation~\cite{Henderson:1995}. All approaches to structure determination are primarily dedicated to overcoming, or working around, the effects of this radiation damage. In X-ray crystallography, large numbers of aligned molecules amplify the diffraction signal that can be obtained within the exposure that the sample can tolerate. The tolerable dose can be increased somewhat by cooling the crystals to cryogenic temperatures. Such cooling also allows electron microscopy---where the ratio of the image-forming to damage-causing radiation is more favourable---to record faint and noisy images of many uncrystallised molecules, which can then be used to build up a three-dimensional image. The extreme intensity and ultrashort pulses of X-ray free electron lasers (XFELs) potentially offer another way to obtain structural information from single macromolecules, but without the need for cooling~\cite{Neutze:2000}. Pulses of femtosecond duration can outrun radiation damage and essentially freeze the molecule in time~\cite{Chapman:2006,Barty:2012}.

Single particle imaging at XFELs consists of collecting coherent diffraction patterns from individual particles intersecting bright XFEL pulses. Theoretical work predicts that currently available XFEL sources generate enough scattered photons from single macromolecules to solve for their unknown orientations and reconstruct 3D structures of large reproducible biomolecules~\cite{Fortmann-Grote:2017,AyyerG:2015,Ayyer:2019}. Proof-of-principle SPI experiments on biological particles~\cite{Bogan:2008,Seibert:2011,Ekeberg:2015,Munke:2016,Reddy:2017,Lundholm:2018,Sobolev:2020} have highlighted the challenges of the approach i.e. the recording of a large number of patterns, all with sufficiently low background, and from structurally homogeneous samples. 

Here, we present experimental results that address these problems and show the path towards single-particle imaging of macromolecules. We addressed the first challenges by aerosol injection of gold nanoparticles and achieved millions of patterns by using the megahertz-rate European XFEL~\cite{Decking:2020} and a relatively large illumination area of the XFEL beam. The particles were chosen for the high scattering power of gold, which balances the reduced intensity from the large beam size to provide scattering signals at the levels expected from biological materials once tight focusing is achieved. 

The challenge of structural heterogeneity is addressed computationally. Even though individual diffraction patterns contained as few as 0.0012 photons per pixel on average, we show that this is sufficient to not only extract the orientations of particles, but also to disentangle structural variations. We obtain a 3D structure approaching \SI{2}{\nano\meter} resolution, which is significantly improved compared to what could be achieved without structural sorting.

With further improvements in aerosol sample delivery to increase the particle density in the X-ray focus~\cite{Roth:2018,Bielecki:2019,Samanta:2020}, more highly focussed X-ray beams can be used to obtain similar data from biomolecules. The computational techniques developed here also open the way to experiments that can reveal thermodynamically rare states in an ensemble and characterise heterogeneous ensembles with statistical rigour. The short exposure times set by the femtosecond pulse duration will also offer unprecedented opportunities for capturing the dynamics of macromolecules in real time.

\section*{Results}
\subsubsection*{Diffraction data collection}
\begin{figure}
  \centering
  \includegraphics[width=0.95\textwidth]{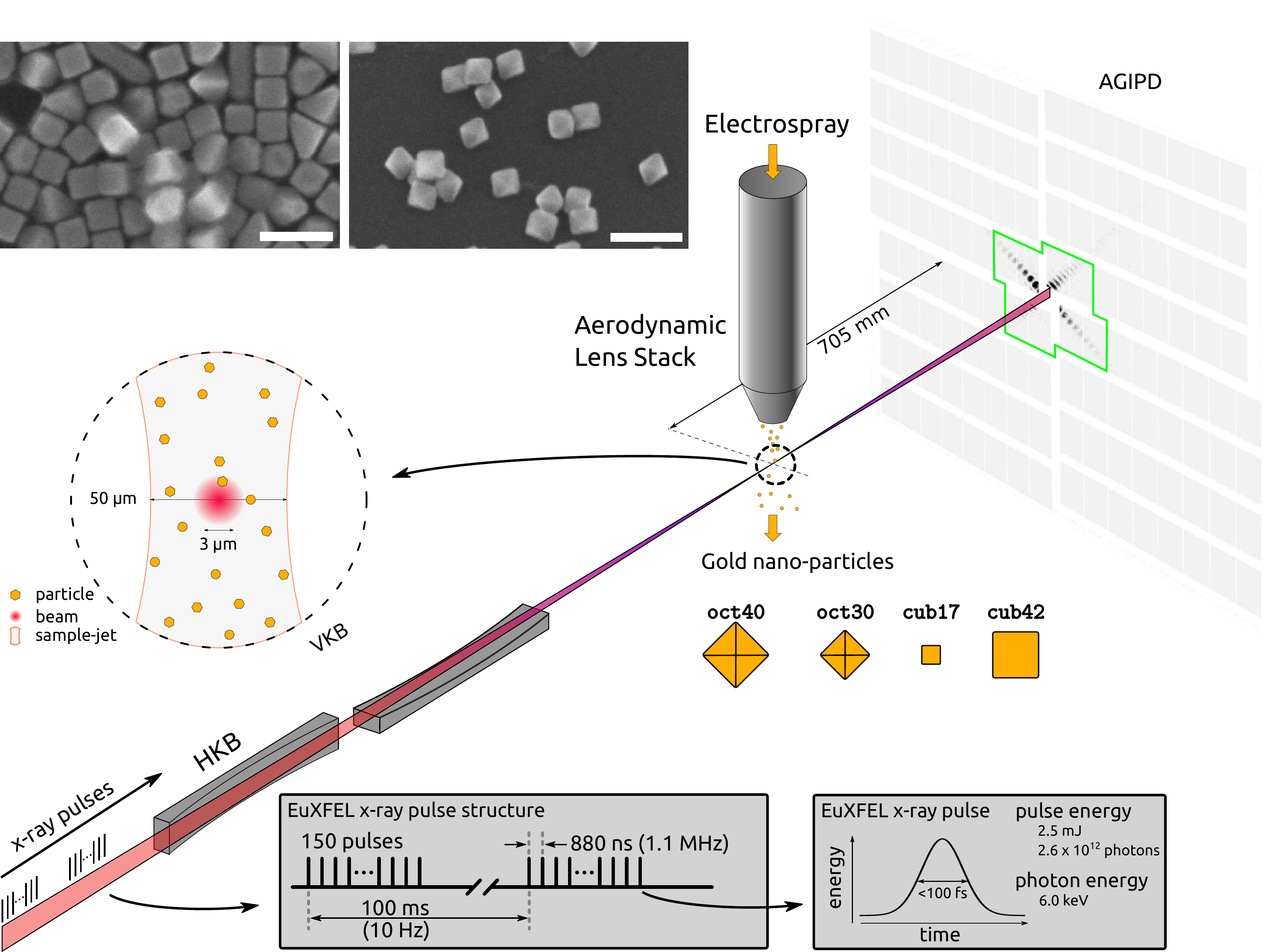}
  \caption{Experimental setup. XFEL pulses were focused by a series of Kirkpatrick-Baez mirrors into a $3\times$\SI{3}{\micro\meter\squared} spot and scattered off particles in the aerosol stream to produce diffraction patterns on the AGIPD. The lower inset shows the timing structure of the XFEL pulses at the instrument while the top inset shows representative SEM images of the \texttt{cub42} and \texttt{oct30} samples; scale bars are \SI{100}{\nano\meter}. The low-resolution part of the detector used for the structural sorting is highlighted in green.}
  \label{fig:setup}
\end{figure}

Data was collected at the SPB/SFX (single particles, biomolecules and clusters/serial femtosecond crystallography) instrument~\cite{Mancuso:2019} of the European XFEL using \SI{6}{\kilo\electronvolt} photons focused into a $3\times$\SI{3}{\micro\meter\squared} spot, as measured by a \SI{20}{\micro\meter}-thick YAG screen in the focal plane. Individual X-ray pulses were generated with \SI{2.5}{\milli\joule} of energy on average (\num{2.6e12} photons). The pulses were delivered in 150-pulse trains with an intra-train repetition rate of \SI{1.1}{\mega\hertz} and trains arriving every \SI{0.1}{\second}, leading to a maximum data collection rate of 1500 frames/second. A detector built specifically for this burst mode operation, the AGIPD~\cite{Heinrich:2011}, was placed \SI{705}{\milli\meter} downstream of the interaction region to collect the diffraction patterns for each pulse individually up to a scattering angle of \SI{8.3}{\degree} at the center-edge of the detector (see Fig.~\ref{fig:setup}).

Gold octahedra and cubes, each of two different sizes, were sequentially injected into the X-ray beam using an electrospray-ionisation aerodynamic-lens-stack sample delivery system (see Methods). The nominal sizes of the particles measured using scanning electron microscopy were 30 and \SI{40}{\nano\meter} for the octahedra and 42 and \SI{17}{\nano\meter} for the cubes. In the rest of the article these samples are described using the codes \texttt{oct30}, \texttt{oct40}, \texttt{cub42} and \texttt{cub17} respectively. The octahedra and cubes were prepared using different protocols, generating different heterogeneity profiles as will be seen later.

Diffraction patterns were observed in around 10~\% of the collected frames. This relatively high hit ratio compared to those achieved with biological particles in similar conditions was due to a combination of the relatively large X-ray focal spot size, high particle concentration and high mass and density of the larger gold nanoparticles, leading to lower speeds after acceleration by the gas flow in the aerodynamic lens stack~\cite{Roth:2018,Awel:2016,Hantke:2018}. Lower speeds lead to higher spatial densities, and thus higher hit ratios for the same particle beam size. Table~\ref{tab:data} shows the statistics of the number of frames collected for each sample as well as the various filtration steps after the analyses described below.

When using the peak repetition rate of \SI{1.1}{\mega\hertz} and 150 pulses per train, diffraction patterns corresponding to the shapes of cubes and octahedra could be observed, but a high fraction of the diffraction patterns appeared to originate from spherical particles (see Table~\ref{tab:data} and third column of Fig.~\ref{fig:2dclass}). This was found to be caused by the melting of particles in the wings of the previous XFEL pulse in the train, as the particles approached the focus. To reduce this occurrence we therefore reduced the intra-train repetition rate from 1.1 MHz to 550 kHz, providing only half the available pulses; further reduction of the repetition rate was tested but not found to be necessary. This reduced-rate mode was used to collect most of the data for the three larger samples (but not the \texttt{cub17} sample).

\begin{table}[h]
\begin{center}
\begin{spacing}{1.9}
\begin{threeparttable}
    \caption{\linespread{1}\selectfont Data collection statistics for the four nanocrystal samples. The sample names refer to their nominal shape (octahedron or cube) and edge length in \si{\nano\meter}.}
    \begin{tabular}{l l l l l}
        \textbf{Parameter} & \texttt{oct30} & \texttt{oct40} & \texttt{cub42} & \texttt{cub17} \\
        \hline
        No. frames & \num{15805472} & \num{29309832} & \num{34197950} & \num{36966286} \\
        No. hits & \num{2117732} & \num{2133041} & \num{2451068} & \num{3307723} \\
        Hit ratio & 13.40\% & 7.28\% & 7.17\% & 8.95\% \\
        Hits/hour & \num{376947} & \num{233553} & \num{228633} & \num{402954} \\
        Hits/train\tnote{*} & 5.2/10.4/15.6 & 2.8/6.4/8.4 & 2.4/5.6/9.1 & NA/7.2/12.1 \\
        No. `good' hits & \num{1430086} & \num{1249328} & \num{433259} & \num{564121} \\
        Sphere fraction (\%)\tnote{*} & 3.4/4.0/19.2 & 2.7/7.2/33.5 & 2.4/10.4/29.1 & NA\tnote{$\dagger$} \\
        Resolution (nm)\tnote{$\ddagger$} & 3.50 (2.10-4.54) & 5.32 (1.89-7.17) & 4.89 (1.98-6.56) & 2.11 (1.81-3.31) \\
        \hline
    \end{tabular}
    \begin{tablenotes}
        \linespread{1}\small
        \item[*] The three numbers correspond to values for 0.28 MHz, 0.55 MHz and 1.1 MHz intra-train repetition rates respectively
        \item[$\dagger$] There was no clear sign of spherical particles for \texttt{cub17} sample
        \item[$\ddagger$] The first number is the azimuthal average resolution while numbers in parentheses show minimum and maximum values, respectively
    \end{tablenotes}
    \label{tab:data}
\end{threeparttable}
\end{spacing}
\end{center}
\end{table}

\subsubsection*{Single hit selection by 2D classification}
Frames with diffraction from particles were detected by setting a threshold on the number of pixels in the AGIPD detector that recorded at least one photon (see Methods). Unfortunately, not all the particles are of interest, even accounting for the heterogeneity. The extraneous patterns include those from spheres formed after melting, multi-particle aggregates and other possible contaminants. In previous work, either manual selection~\cite{Ekeberg:2015,Lundholm:2018} or manifold learning methods~\cite{Yoon:2011,Rose:2018,Reddy:2017} have been used to classify patterns and reject outliers. We adopt an alternative approach, similar to one commonly used in cryo-EM~\cite{Scheres:2005}, but implemented in diffraction space. Two-dimensional orientation determination into multiple models was performed in the detector plane using the EMC algorithm~\cite{Loh:2009,Loh:2010} implemented in \textit{Dragonfly}~\cite{Ayyer:2016}. The in-plane rotation angle ($\theta$) and relative incident fluence ($\phi$) of each diffraction pattern was determined collectively and multiple independent 2D intensity models were reconstructed. Each of these intensities represent an average of aligned copies of a subset of the patterns from the whole set. In addition to the EMC algorithm being highly noise-tolerant~\cite{Philipp:2012,Giewekemeyer:2019,Ayyer:2019}, one can also use it to examine the average models to understand what type of particles are in the dataset.

\begin{figure}
\centering
\begin{tabular}{ c >{\centering\arraybackslash}m{0.86\textwidth} }
    \rotatebox[origin=c]{90}{\texttt{oct30}} & \includegraphics[width=0.85\textwidth]{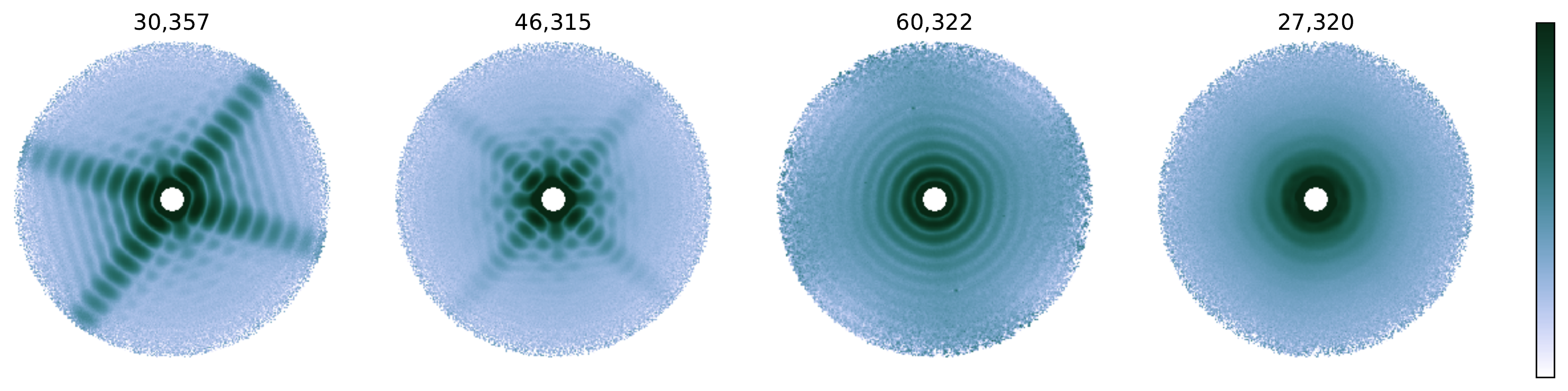} \\
    \rotatebox[origin=c]{90}{\texttt{oct40}} & \includegraphics[width=0.85\textwidth]{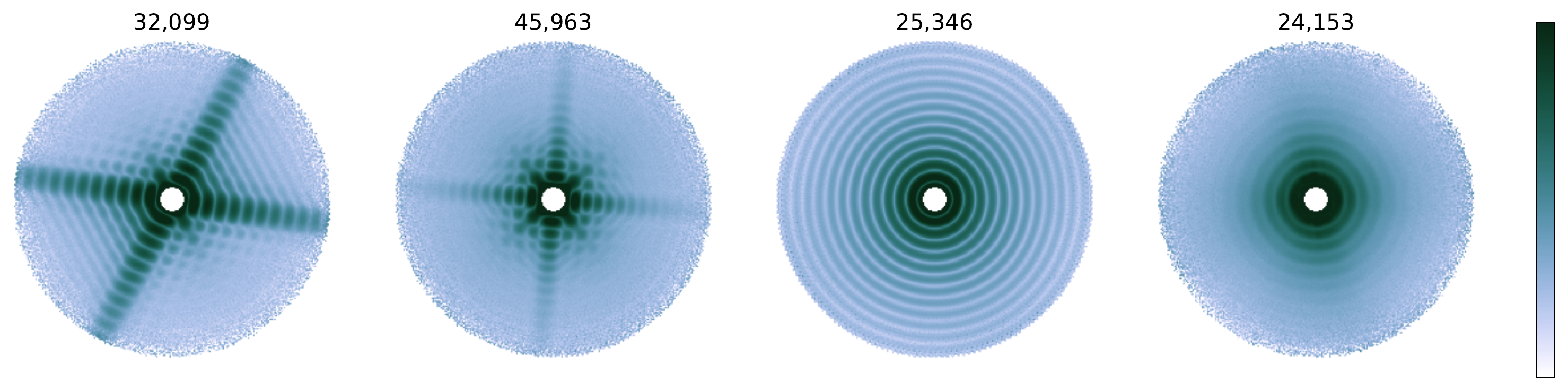} \\
    \rotatebox[origin=c]{90}{\texttt{cub42}} & \includegraphics[width=0.85\textwidth]{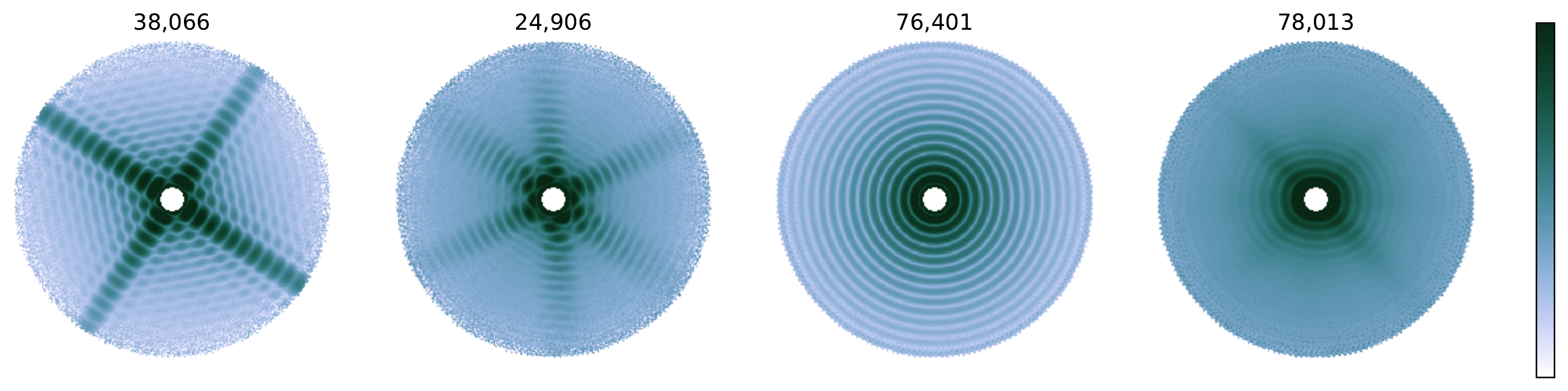} \\
    \rotatebox[origin=c]{90}{\texttt{cub17}} & \includegraphics[width=0.85\textwidth]{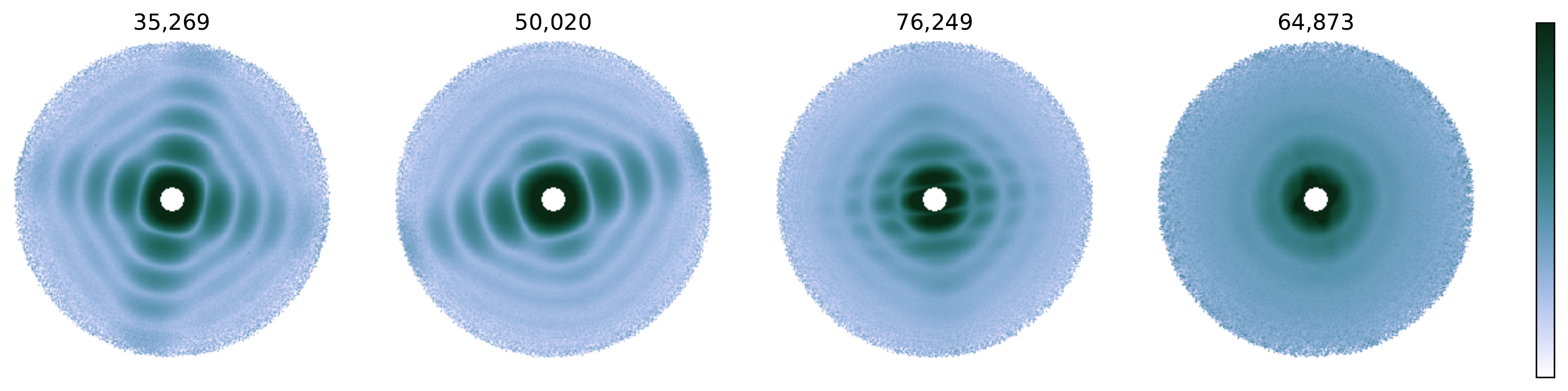} \\
\end{tabular}
\caption{Representative examples of reconstructed 2D models shown on a logarithmic scale, with each row representing a different sample. The numbers indicate how many patterns had that model as the most likely one. The first two columns show models selected for further processing. The third column shows diffraction from rounded/spherical particles, except in the \texttt{cub17} case where there were no spherical particles and the model shows diffraction from a dimer instead. The fourth column shows some of the low-contrast models generated by averaging patterns from a diverse set of particles. The resolution at the edge of the circle is \SI{3.3}{\nano\meter}.}
\label{fig:2dclass}
\end{figure}

In this experiment, 50 random white noise 2D intensity models were used as initial guesses to perform the classification for each sample, using only the low resolution part of the detector highlighted at this stage (see Fig.~\ref{fig:setup}). Some of the reconstructed intensities are shown in Figure~\ref{fig:2dclass}. The first two columns of the figure show representative examples of `good' models of each sample, chosen manually to be those with high contrast and strong streaks for further processing. The third column shows an average of diffraction from rounded particles (except in the \texttt{cub17} case where a dimer average is highlighted). These models were used to determine the sphere fraction shown in Table~\ref{tab:data}. Finally, the last column shows low-contrast models where a diverse set of particles were averaged.

\begin{figure}
\centering
% PDF versions of 2D histograms are there in the folder, but they slow down compilation
\begin{tabular}{ c c }
    \includegraphics[width=0.45\textwidth]{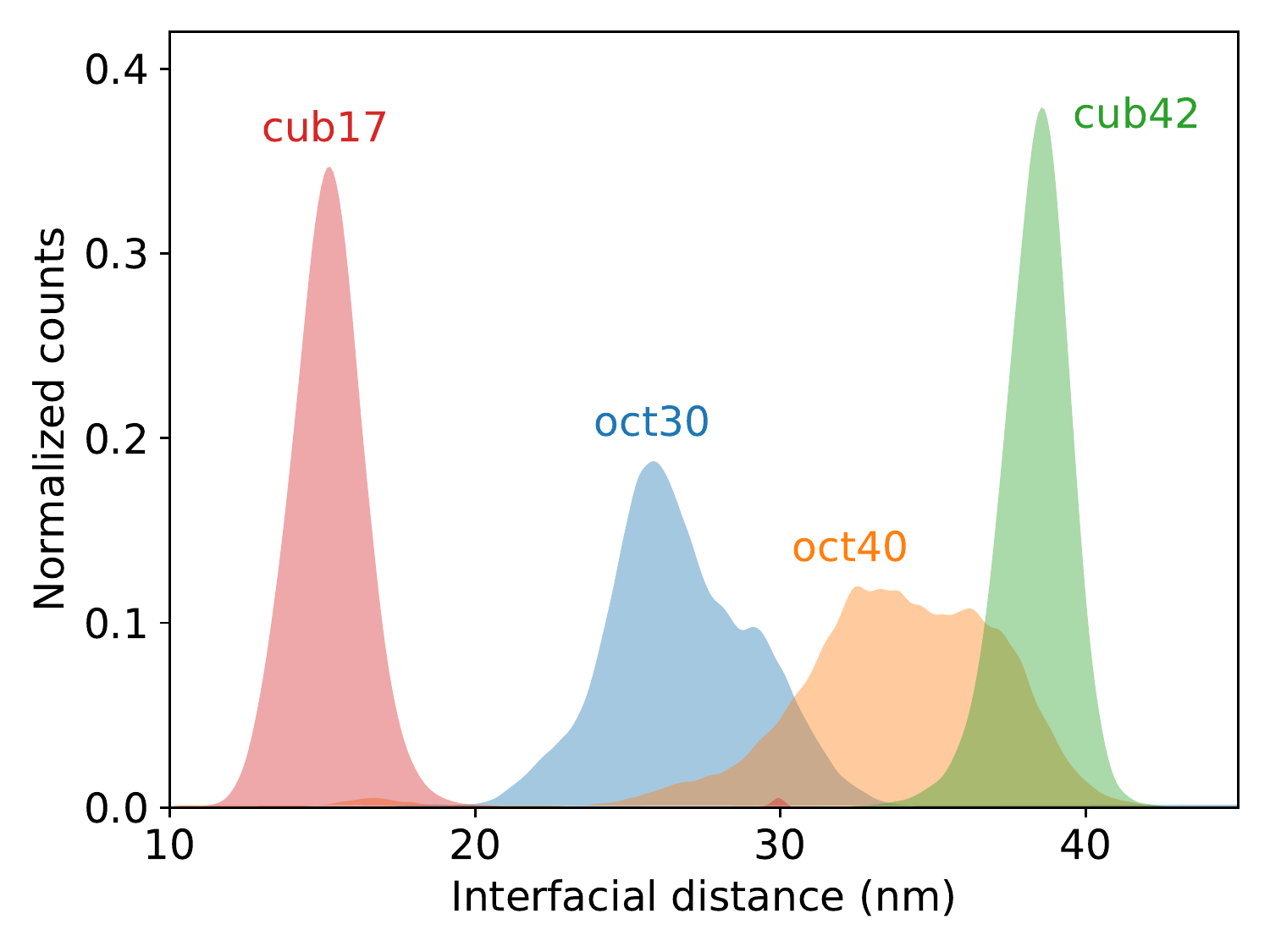} & \includegraphics[width=0.45\textwidth]{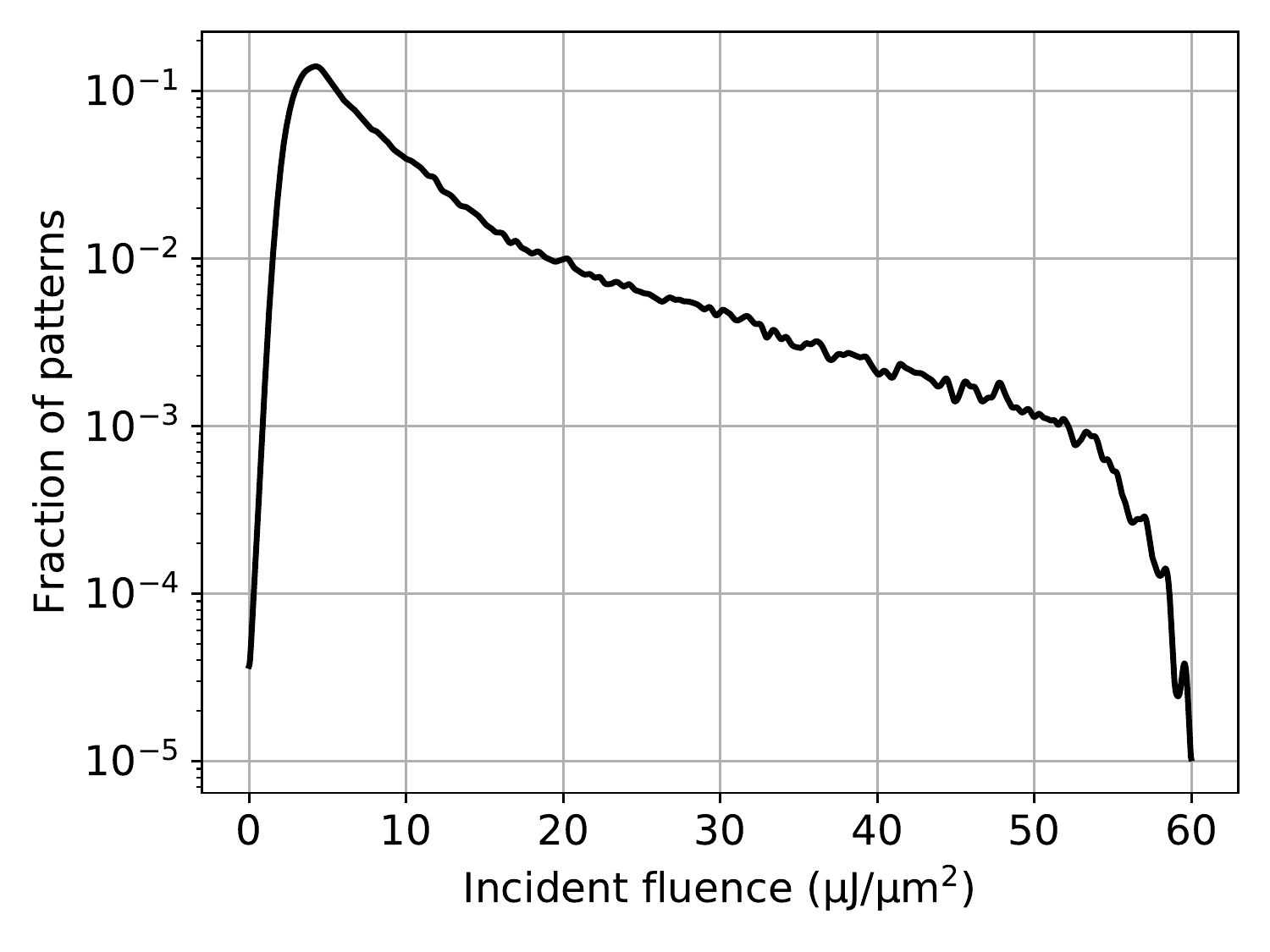} \\
    (a) & (b) \\
    \includegraphics[width=0.45\textwidth]{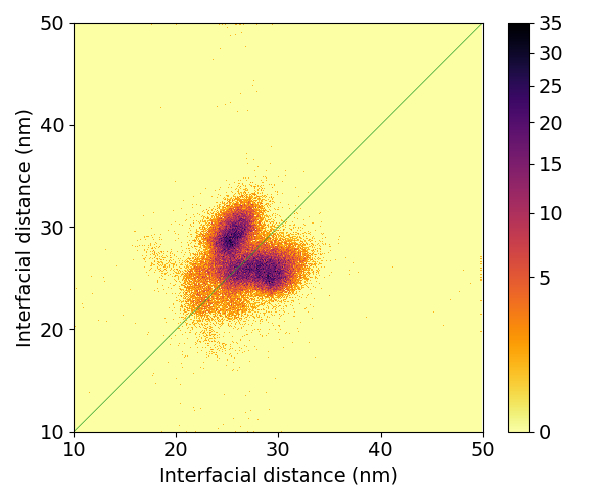} & \includegraphics[width=0.45\textwidth]{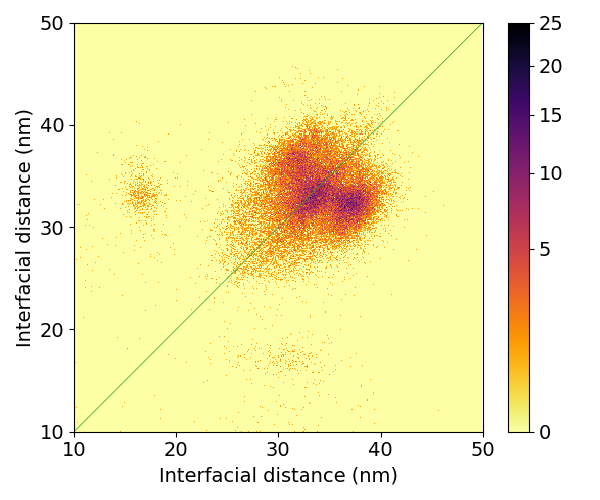} \\
    (c) & (d)
\end{tabular}
\caption{Size and incident fluence distributions from 2D classification. (a) Size distribution for the 4 samples. The sizes are represented by the distance between opposing parallel faces. The cubes have narrow distributions, while the octahedral distributions are broader. (b) Distribution of incident fluence on the particle calculated from the \texttt{cub42} sample assuming they are ideal cubes. (c-d) 2D histogram of size distributions from two-streak patterns for the \texttt{oct30} and \texttt{oct40} samples respectively. High density in the off-diagonal regions suggests the particles were asymmetric. The horizontal axis represents the brighter of the two streaks.}
\label{fig:sizing}
\end{figure}

The 2D classification also enabled the analysis of size-heterogeneity from those models where the faces of the nanoparticles were parallel to the X-ray beam. In these cases, one observes strong streaks on the detector and the fringe spacing indicates the distance between these parallel faces. The size distributions of the samples inferred this way are shown in Fig.~\ref{fig:sizing}(a). The octahedral samples had a much broader size distribution than the cubic ones. While some of the breadth of the peaks is due to apparent size variations when the faces are not being perfectly parallel to the beam, the much broader size distributions of the octahedra suggest that they had more heterogeneity. 

In addition, the octahedra were also noticeably asymmetric, as seen in Figs.~\ref{fig:sizing}(c) and (d). These histograms were made by identifying patterns which belonged to models with two strong streaks (e.g. top left model in Fig.~\ref{fig:2dclass}). Another run of 2D classification with just these two-streak patterns showed no variation in the angle between the streaks, but only in the fringe spacing. This is to be expected since the angle is fixed by the $\langle 111 \rangle$ growth direction, while the size is not restricted by symmetry. The equivalent figures for the cubic samples showed no asymmetry. 

Due to the low polydispersity of the cubes, they were used to determine the incident fluence distribution of the X-ray beam. Since the Fourier transform of a cube is the product of three orthogonal \textit{sinc} functions, the size fitting procedure also generated a predicted incident fluence. The distribution from \num{102480} patterns is shown in Fig.~\ref{fig:sizing}(b), yielding a maximum fluence of around \SI{60}{\micro\joule/\micro\meter\squared}, which leads to a lower bound estimate of around \SI{540}{\micro\joule} in the focal spot from the measured spot size. The actual fluence was likely higher as the particles were not ideal cubes and the scattering efficiency is reduced at high fluences~\cite{Jonsson:2015,Ho:2020}. One can also see that most diffraction patterns were obtained with lower incident fluences, because the particles interacted with the outer regions of the X-ray focus.

\subsubsection*{3D reconstruction with structural sorting}
The fraction of good hits used for 3D structure reconstruction varied from 17~\% for the cube samples to around 60~\% for the octahedra (see Table~\ref{tab:data}). The 3D intensity distribution was obtained using these patterns before recovering the structures by performing phase retrieval using the difference map algorithm~\cite{Elser:2003,Ayyer:2019}. For computational efficiency, the 3D orientations were first determined using the low-resolution part of the detector where the highest resolution was \SI{3.3}{\nano\meter}. A refinement procedure similar to that developed for serial crystallography~\cite{Lan:2017} was used with the whole detector to get the full-resolution 3D intensities. In this procedure, only orientations in the neighbourhood of the most likely orientation of a given pattern from the low-resolution run were searched.

The intensities recovered in this manner had noticeably lower contrast than the equivalent slices in the 2D models. From the size distributions seen in Fig.~\ref{fig:sizing}, this could be attributed to structural heterogeneity. To counter this, the patterns were probabilistically partitioned into five intensity volumes in a manner equivalent to the 2D classification procedure. However, the initial guesses were not random white noise, but rather isotropically stretched/scaled versions of the average models reconstructed above. Five models, with stretch factors ranging from 0.9 to 1.1 were used as these initial seeds. The rest of the reconstruction proceeded without any restraints between these models or any symmetry constraints.

\begin{figure}
\centering
% Will switch to PDF version of sizehist at the end
\begin{tabular}{ c c }
    \multirow{3}{*}[0.27\textwidth]{\includegraphics[width=0.65\textwidth]{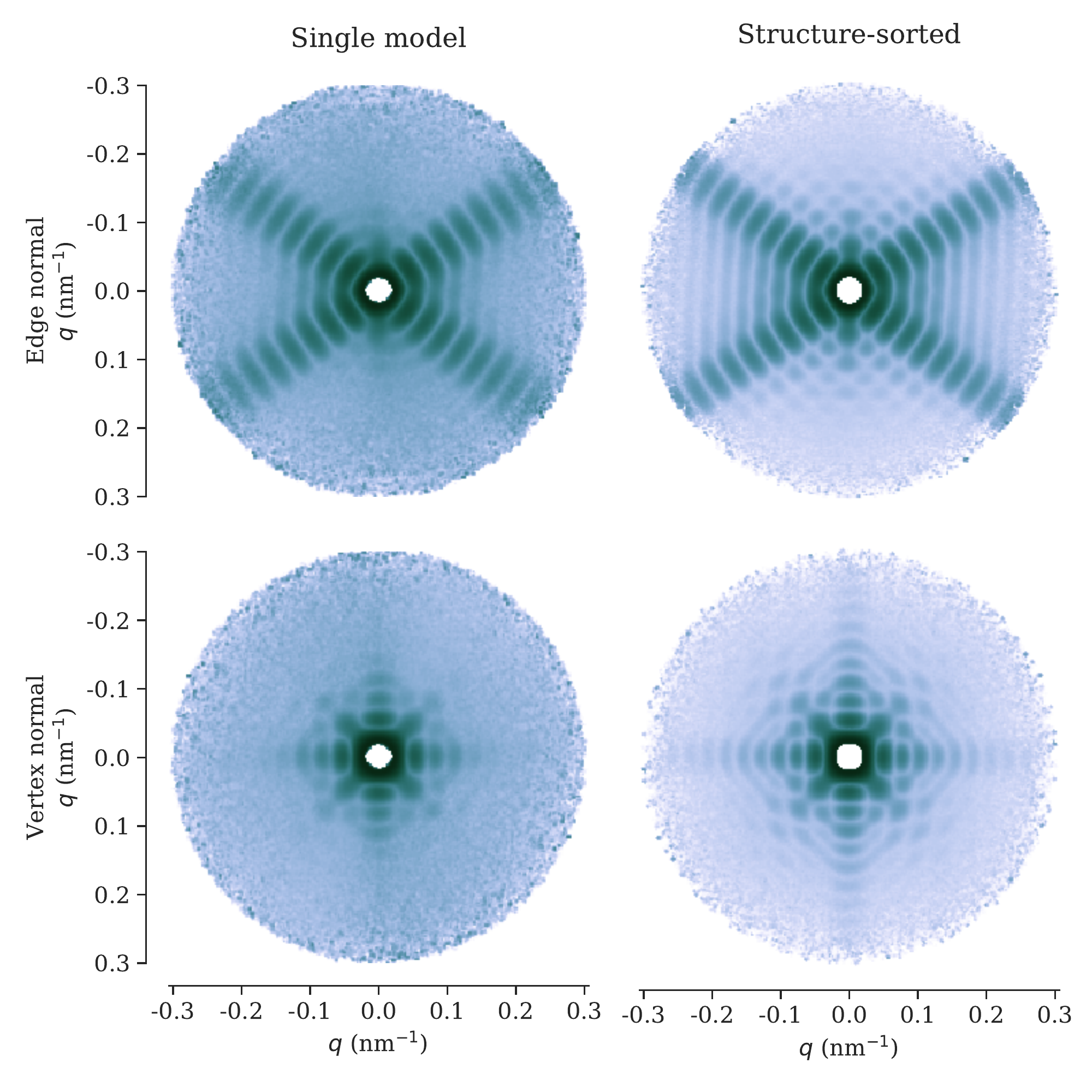}} & \includegraphics[height=0.3\textwidth]{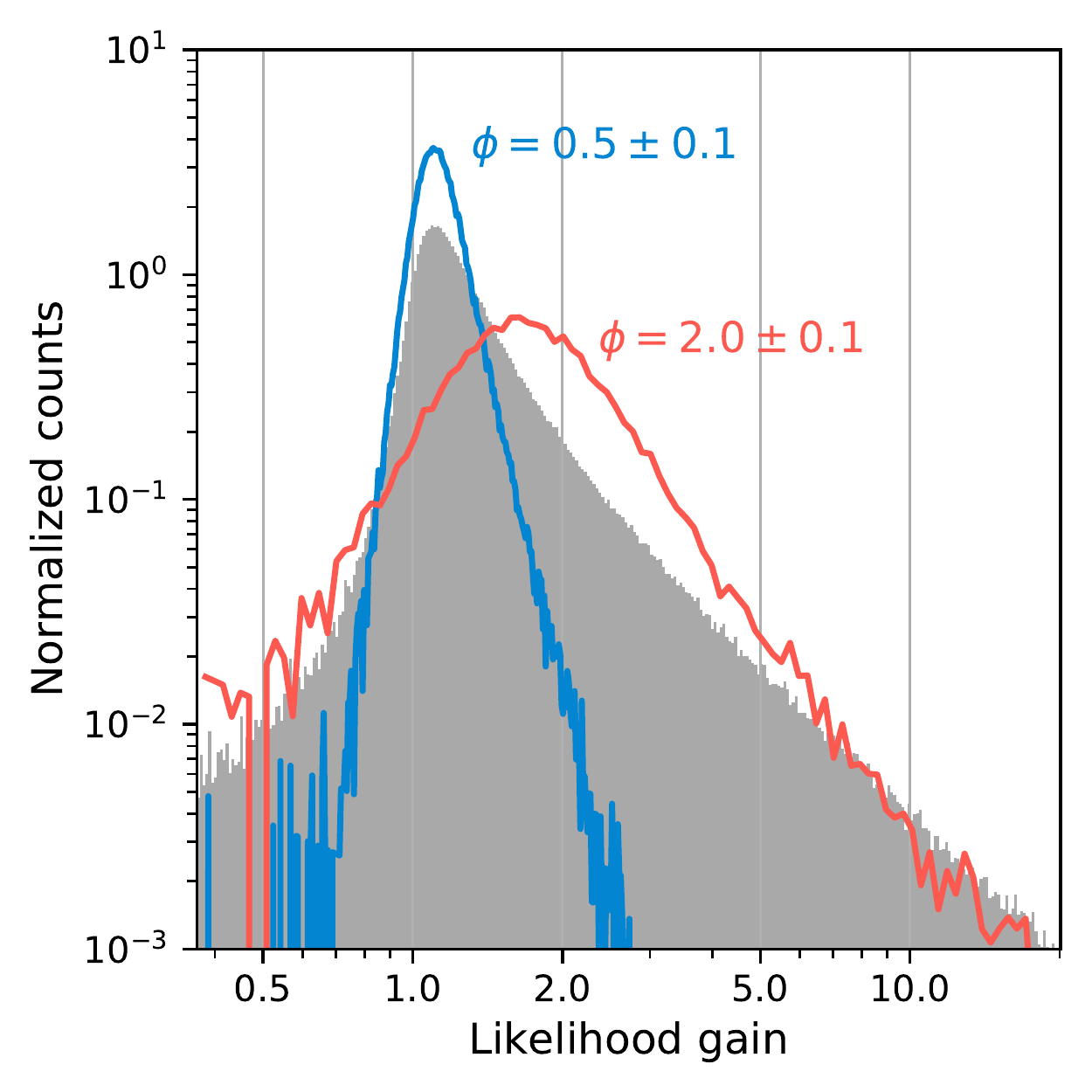} \\
     & (b) \\
     & \includegraphics[height=0.3\textwidth]{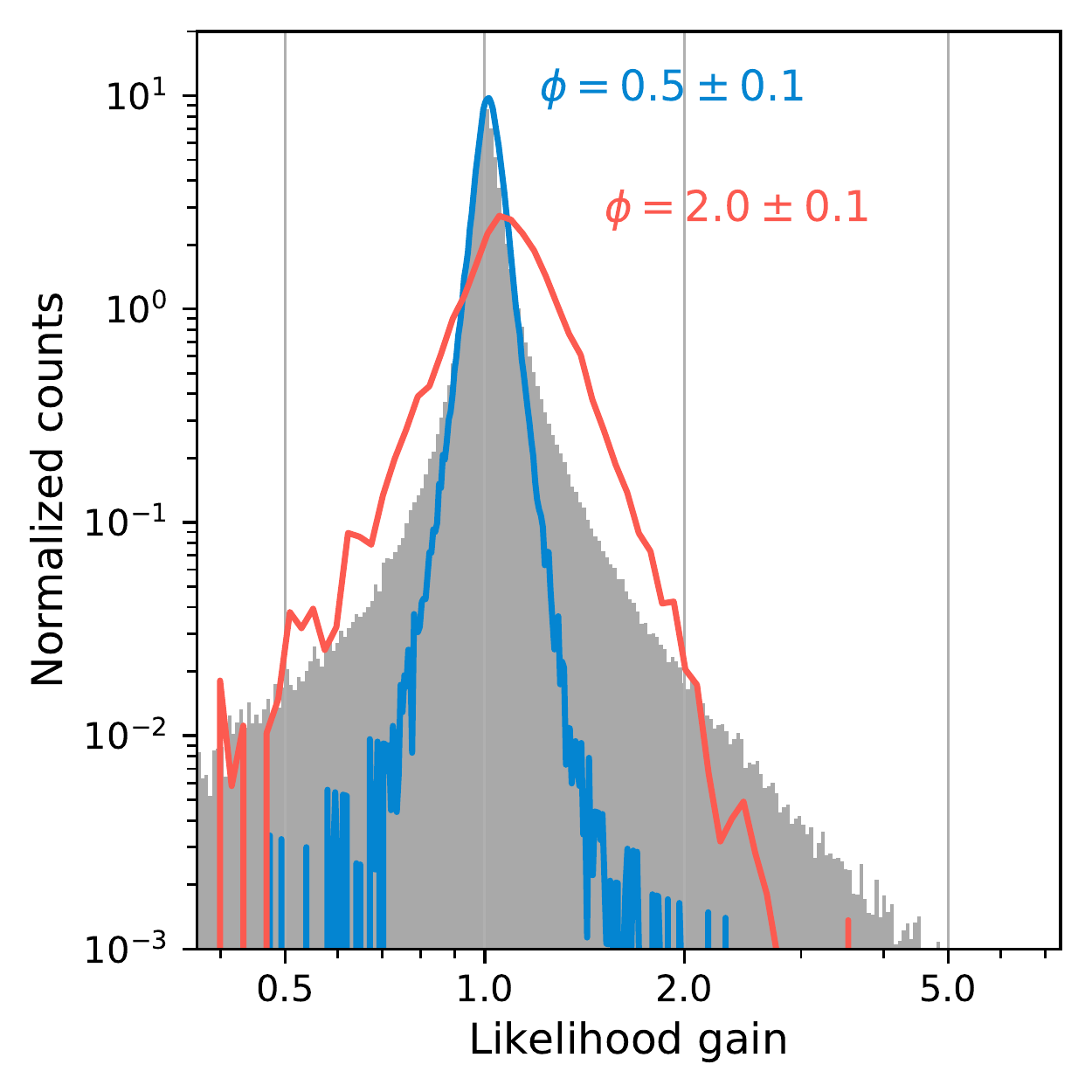} \\
    (a) & (c) \\
    \multicolumn{2}{c}{\includegraphics[width=0.95\textwidth]{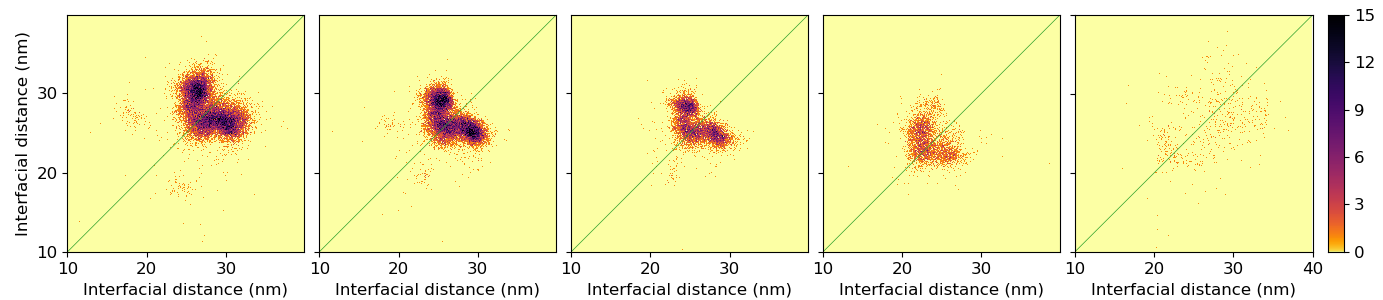}} \\
    \multicolumn{2}{c}{(d)}
\end{tabular}
\caption{Comparison of 3D intensity reconstructions for the octahedra before and after structural sorting. (a) Low-resolution logarithmic intensities of the \texttt{oct30} sample comparing the standard single-model reconstruction with one of the sorted models. The two rows represent slices normal to an edge and vertex of the octahedron respectively. (b) Likelihood gain distribution for the patterns which are shared with the sorted model shown in (a). The blue and red curves show distributions for weak and strong patterns, as identified by the relative fluence factor $\phi$, respectively. (c) The same gain plot for the \texttt{oct40} sample. (d) Two-streak size histograms (see Fig.~\ref{fig:sizing}(c)) for the \texttt{oct30} sample separated into the five reconstructed models.}
\label{fig:3dintens}
\end{figure}

Once again, this structural sorting was performed at low resolution before refining the orientations of a subset of patterns from a single model to get full-resolution intensities. A comparison of orthogonal slices through the 3D intensity for the \texttt{oct30} sample is shown in Fig.~\ref{fig:3dintens}(a). The left column, showing the single model reconstruction with 1.4 million patterns has noticeably worse fringe contrast and background than the equivalent slices in the right column or in the first two columns of the 2D classification output shown in Fig.~\ref{fig:2dclass}. The homogeneous set had 0.53 million patterns selected using the multi-model EMC reconstruction. The visual improvement is accompanied by an increase in the likelihood of the model intensities outside the central speckle for the common patterns in both sets, as shown in Fig.~\ref{fig:3dintens}(b). The filled histogram shows the distribution of the per-pattern increase in likelihood, which we refer to as likelihood gain, while the two traces show the distributions for weak (relative scale $0.5 \pm 0.1$) and strong (relative scale $2.0 \pm 0.1$) patterns. The latter shows how brighter patterns are more selective towards an improved model. Figure~\ref{fig:3dintens}(c) shows the same information for the \texttt{oct40} sample, where the gain ratio is smaller, but still greater than 1. The 2D size distributions shown in Fig.~\ref{fig:sizing}(c) were re-calculated for each subset of patterns belonging to the five models and plotted in Fig.~\ref{fig:3dintens}(d), confirming the different sizes for each model, but also exhibiting a simpler structure than that of the full dataset.

For the cubic particles, a single model 3D reconstruction was deemed sufficient, due to the relative monodispersity of the sample. The selection of `good' hits from the 2D classification was more stringent, including only high-contrast cube-like patterns. The incident fluence factors were estimated in the first few iterations where the calculated probability distributions were broad and then later kept fixed (see Methods).

\begin{figure}
\centering
\begin{tabular}{c c}
    \includegraphics[width=0.45\textwidth]{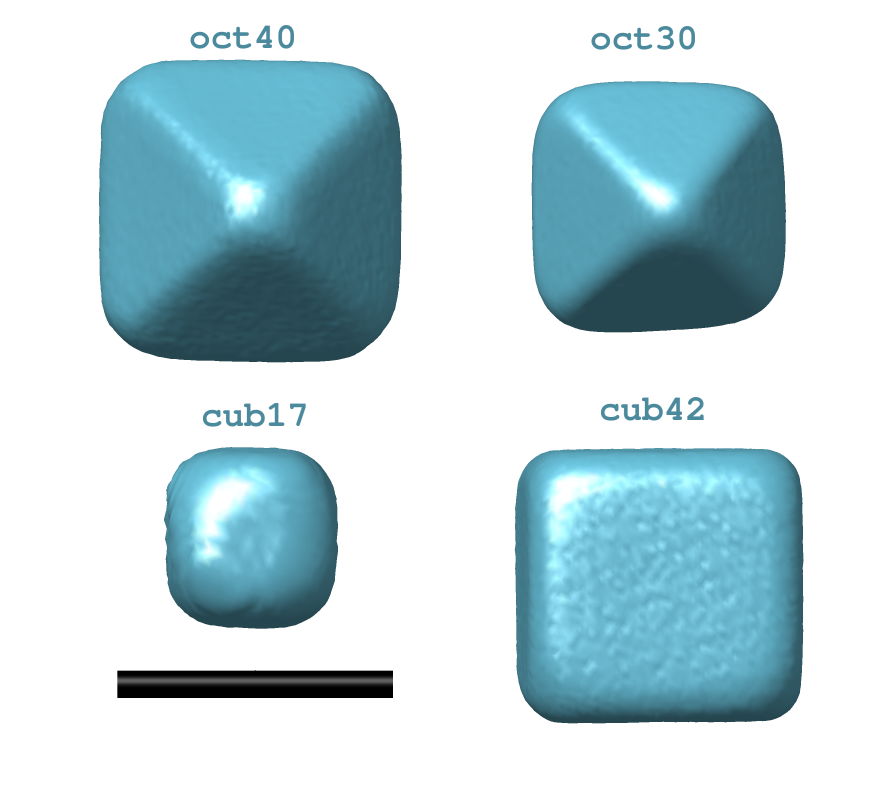} & \includegraphics[width=0.45\textwidth]{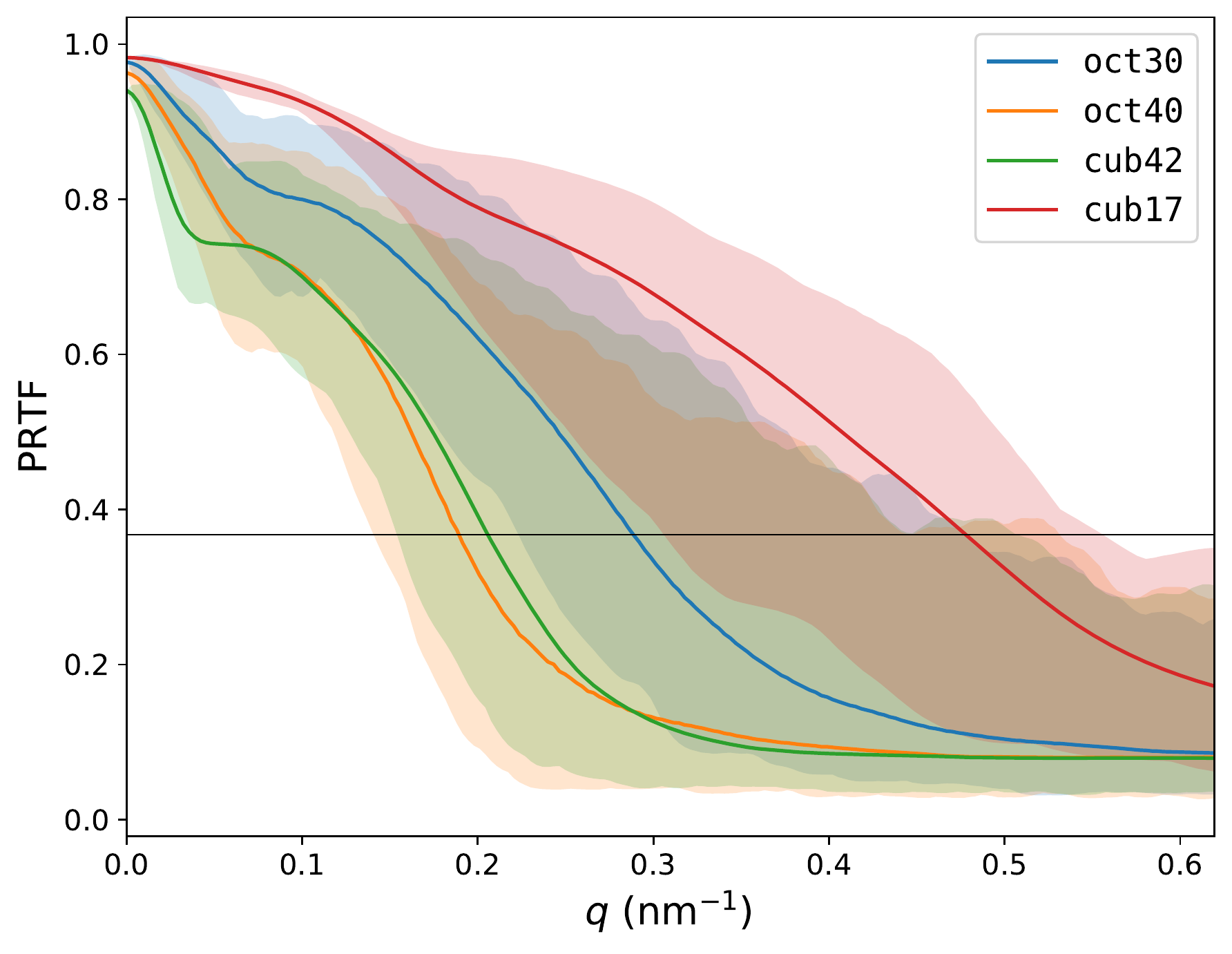} \\
    (a) & (b) \\
\end{tabular}
\caption{Phase retrieval. (a) Isosurface plots of electron densities recovered after phase retrieval (scale bar is \SI{40}{\nano\meter}). The asymmetric structures of the octahedra are clearly evident (see Supplementary Movie S2). (b) Smoothed phase retrieval transfer function (PRTF) measuring reproducibility of phases as a function of $q$. The solid lines represent the azimuthal average PRTF conventionally used to determine the resolution of the structure. The shaded region around each line indicates the range of values at each $q$. The typical $1/e$ cutoff is shown in black.}
\label{fig:phasing}
\end{figure}

The electron densities were reconstructed by performing 3D iterative phase retrieval on the full-resolution intensity volumes (see Methods for details and Supplementary Fig.~\ref{fig:intensa} for intensity slices). Figure~\ref{fig:phasing}(a) shows the reconstructed electron densities as isosurface plots. The contour levels were chosen where the gradient of the density was highest. The phase retrieval transfer function (PRTF) metric as a function of wavevector $q$ is shown in Fig.~\ref{fig:phasing}(b). This metric is a measure of the reproducibility of recovered phases when starting from 128 random models. The 3D PRTF distribution was smoothed using a Gaussian kernel with a width equal to 1/3rd of the fringe width. The shaded region around each line shows the range of values in each $q$ shell, highlighting the strong anisotropy of the metric due to the faceted nature of the objects. The intersection with the common $1/e$ threshold determining the resolution is shown in Table~\ref{tab:data}. The resolution normal to the flat faces is \SI{2}{\nano\meter} or better for all samples, while the resolution is relatively low far from any strong streaks in Fourier space. This angle-dependent resolution is a property of the diffractive-domain averaging before phase retrieval, but also due to the strongly faceted shape and lack of internal structure of these objects, both of which are not representative of biological objects.

\section*{Discussion}
We have demonstrated an order-of-magnitude increase in data collection efficiency along with much higher imaging resolution than previously achieved for X-ray single particle diffractive imaging, setting a template for future SPI experiments at the European XFEL and elsewhere. We have also shown that with these large data sets, one can structurally sort the particles and average a narrow size and shape range to obtain higher resolution. A similar problem is expected to be faced when imaging biological particles and the method developed here shows the way towards overcoming conformational variability in the Fourier domain.

Although we benefited from the strong scattering cross section of gold compared to organic materials, with the commissioning of a sub-micron focus at the SPB/SFX instrument, we can expect comparable signal strengths from organic materials. Unfortunately, smaller X-ray foci would also mean lower hit ratios with the current sample delivery setup. Improvements could be made through optimised focussing for the targeted size distribution~\cite{Roth:2018} or cryogenic injections systems~\cite{Samanta:2020} which additionally allow conformational selection~\cite{Chang:2015}. Another approach is to keep using the larger focus and conjugate the particles with gold nanoparticles to assist hitfinding and orientation determination~\cite{Ayyer:2020}. The effective hit rate can also be increased by using more pulses from the European XFEL (max. 2700) than the AGIPD detector can save (max. 352) and vetoing in real time those frames which do not contain diffraction signal. 

The class of experiments exemplified here can also be applied to study rare events such as transient states in a spontaneous phase transition or high free-energy states. Since each image is collected serially, one can identify relevant subsets corresponding to interesting states without averaging over all patterns. In this work, we have taken the approach of treating the objects as general 3D contrast functions with no \textit{a priori} information. One can also envision a parameterised refinement approach which should enable a finer characterisation of the structural landscape of the ensemble.

\section*{Methods}
\subsubsection*{Sample preparation}
The octahedral gold nanoparticles were synthesised using published protocols~\cite{Li:2008,Lu:2017} with poly(diallyl\-dimethyl\-ammonium) chloride (PDDA) polymer coating to avoid aggregation. The cubic particles were synthesized in water using the method described in Park~\emph{et~al.}~\cite{Park:2018} with cetyltrimethylammonium chloride (CTAC) as a stabilising agent. In order to obtain the requisite $10^{12} - 10^{13}$ particles/\si{\milli\liter} concentration to approach an average of one particle per electrospray droplet~\cite{Bielecki:2019}, all syntheses were concentrated from initial values of $10^9$~particles/\si{\milli\liter} for the cubes and $10^{11}$~particles/\si{\milli\liter} for the octahedra and excess ligands were removed by centrifugation. Scanning and transmission electron microscopy images of the samples are shown in Supplementary Figs.~\ref{fig:sem_sample} and \ref{fig:tem_sample}.

\subsubsection*{Aerosol sample delivery}
The samples were suspended in 10 mM ammonium acetate and aerosolized using an electrospray nebulizer (average flow rate \SI{200}{\nano\liter/\minute}) and neutralized before delivering into the X-ray interaction point using the aerodynamic lens stack~\cite{Bielecki:2019}. An electrospray differential mobility analysis (ES-DMA) setup was installed at the beamline and the particles generated by the electrospray could be diverted into the ES-DMA to characterise the size distribution and concentration of aerosolized particles. Particle size distribution measurements were carried out with an electrostatic classifier (TSI 3082) together with the DMA (TSI 3081). The DMA was connected to a condensation particle counter (CPC, TSI 3789). Representative size distributions are shown in Supplementary Fig.~\ref{fig:dma_sizing}. To diagnose the width and density of the particle stream in the X-ray interaction region after aerodynamic focusing, we employed a Rayleigh scattering diagnostic (see Supplementary Fig.~\ref{fig:rayleigh}) using a frequency-doubled Nd:YAG laser which was mirror-incoupled perpendicular to both the X-ray beam and particle stream~\cite{Awel:2016,Hantke:2018}. These two diagnostic tools helped in assessing both the quality of the samples as well as the transmission efficiency of the sample delivery system.

\subsubsection*{Online monitoring}
In order to help align the experiment and dianose problems during data collection, the \textit{Hummingbird} software~\cite{Daurer:2016} was connected to the \textit{Karabo} bridge in the European XFEL DAQ system~\cite{Heisen:2013} to receive data with a delay of a few seconds. Since most of the photons in an SPI experiment are concentrated at low resolution, only the module of the AGIPD closest to the beam centre was used for online analysis. The use of uncalibrated data and only a single module enabled a frame rate of up to \SI{800}{\hertz} using all 176 memory cells of each pixel of the AGIPD available in this experiment. The analyses conducted live included lit-pixel hit finding and hit-ratio determination (see the following Preliminary analysis section), sphere model size determination and the detection of the fraction of spherical particles by analysing the azimuthal variation in intensities. The latter was used to understand and fix the particle melting issue mentioned in the Results section.

\subsubsection*{Preliminary analysis}
The AGIPD detector was calibrated using offset constants for each cell in each pixel. Except for a few pixels near the beam centre, no pixels switched gain mode. After offset correction, the number of pixels containing at least 0.7 of a photon was calculated for each frame. In the absence of particles, the number of such pixels is normally distributed with a mean dependent on background from the beamline, carrier gas and detector false positives. A threshold of 3$\sigma$ over the mean number in each run was used to select frames with particle scattering. Over the entire experiment, the hit ratio fluctuated between 7\%-15\%. For this and future analyses, memory cells in pixels with outlier dark offsets or dark noise were masked out, along with the double-wide pixels along ASIC edges.

These hits were converted to photons by first subtracting the dark offsets, correcting for per frame common mode shifts by subtracting the median of each 64x64 pixel ASIC and then subtracting a pixel-wise running median of the last 128 frames over all cells. The last step was important in removing artifacts due to the slow drift of dark offsets on a pixel level. These corrected detector values were then converted to integer photon counts by thresholding with a variable cutoff using the following procedure.

The probability distribution of detector ADUs (analog-to-digital units) at a pixel in the absence of photons is a Gaussian centered at 0 with a cell-dependent width. The 1-photon distribution is a shifted copy of the 0-photon distribution with a height which depends on the signal level (ignoring charge sharing for the large \SI{200}{\micro\meter} pixels). The optimal threshold was chosen to be the point at which these 0-photon and 1-photon distributions intersect for a signal level of $10^{-3}$ photons/pixel. If the 1-photon distribution is centered at $m_1$ ADUs and the standard deviation of the noise of a cell is $\sigma$ ADUs, the threshold was
\[t = m_1 \left(0.5 - \frac{\sigma^2}{m_1^2} \log(10^{-3})\right)\]
This threshold minimises the total error rate (false positive plus false negative) due to detector noise at the chosen signal level, ignoring charge sharing effects which are small for the \SI{200}{\micro\meter} pixels of the AGIPD. For higher signals, at lower resolution, the error rate would be higher, but biased towards false negatives. See Supplementary Fig.\ref{fig:detcorr} for the effects of the various corrections on the integrated detector image. For the current detector configuration and photon energy the 1-photon peak was centered at 47 ADUs and the average threshold was 0.755 of a photon.

\subsubsection*{Intensity reconstruction}
Both two- and three-dimensional intensity volumes were reconstructed using the \textit{Dragonfly} package. Detector files were generated and refined manually starting from initial geometries from a previous serial crystallography experiment~\cite{Yefanov:2019}. The geometry refinement only involved adjusting the positions of the detector quadrants since the modules within a quadrant had not been moved between the experiments. Two detector files were produced, one for the inner eight 128x64 pixel detector ASICs, while a high-resolution version contained all 1024x1024 pixels.

The photon-converted data was saved in the sparse \textit{Dragonfly} \texttt{.emc} format with file sizes of around \SI{10}{\giga\byte} per sample. For the 3D reconstructions, the detector files specify the reciprocal space voxel coordinate of each pixel, which involves defining the radius of curvature of the Ewald sphere in voxels. The natural choice of the detector distance in pixel units (3525) produced too large an oversampling factor, with a fringe spacing of around 20 voxels for the largest \texttt{cub42} sample and even higher for the smaller samples. For the low-resolution detector file, the radius of curvature was set to be 2000 voxels, generating $253^3$ voxel volumes. For the full-resolution detector, it was 1500 voxels for all samples except \texttt{cub17}, where it was 1000 voxels. 

For all reconstructions, the deterministic annealing procedure~\cite{Ueda:1998,Ayyer:2016} was used to improve the convergence of the algorithm. The annealing parameter $\beta_d$ was initially set for each pattern, $d$ based on the number of scattered photons using the following empirical formula: 
\[\beta_d = \exp(-1.156\; C_d^{0.15})\]
where $C_d$ refers to the number of orientationally relevant photons in the pattern. This generates a lower value for brighter patterns, broadening their otherwise sharp probability distribution over orientations. The specific expression was tested in simulations to produce a relatively flat dependence of the mutual information $I(K,\Omega)$~\cite{Ayyer:2016} on the signal strength. The parameter was increased by a factor of 2 every 10 iterations for each pattern.

In the 2D reconstructions, 180 angular samples were chosen in the range from 0 to $2\pi$ for each model. The low resolution 3D reconstructions were performed with an orientational sampling level of 8, which corresponds to \num{25680} samples~\cite{Loh:2009}. The high-resolution refinement went up to a sampling level of 20 (\num{400200} samples). For the 3D multi-model reconstructions, the initial intensities were generated by isotropically stretching the single-mode intensity volume using linear interpolation and the initial fluence factors were also used from that run.

For the cubic samples, the 3D reconstruction pipeline was modified. First, 40 iterations were performed with the initial $\beta_d$ parameters without an annealing schedule. For the larger \texttt{cub42} sample, only the pixels corresponding to the first 3 diffraction fringes (\SI{12.9}{\nano\meter} resolution) were used to determine the orientations and fluence factors. This was to avoid instabilities since most of the Fourier power at higher resolution is concentrated in the streaks normal to the faces and the angle between the streaks is large enough that interference between them poorly constrains the model. Once the low resolution, rotationally blurred intensities were stable, the fluence factors were fixed and the annealing schedule was enabled. The rest of the reconstruction proceeded in a similar manner to the octahedra. Isosurface plots for the three larger particles using the low-resolution data are shown in Supplementary Movie 1.

\subsubsection*{Size and incident fluence fitting}
The 2D classification was first used to identify patterns where strong streaks were visible. These classes can be seen in the first column in Fig.~\ref{fig:2dclass}. A pair of parallel faces on a particle produce a \textit{sinc}-function dependence in the Fourier transform along the face normal. This was used to determine the interfacial distance from each pattern. 

The classification procedure not only allowed us to identify the patterns which have strong streaks, but was also used to determine the angles of these streaks since we knew by what angle the pattern had to be rotated to fit the model. For these selected patterns, the intensity distribution along the streak was calculated by integrating over a 21-pixel wide strip along the streak. The size was determined by cross-correlating the intensity distributions with \textit{sinc} functions for sizes from 10 to \SI{50}{\nano\meter} in \SI{0.1}{\nano\meter} increments. The size was chosen as the one which produced the maximum Pearson correlation coefficient and only those streaks with a coefficient greater than 0.9 were included. Figures~\ref{fig:sizing}(c, d) were generated by only considering patterns with 2 strong streaks.

For the \texttt{cub42} sample, the brightness of the two-streak patterns were used to determine the incident fluence by assuming that they were generated by perfect cubes. A procedure similar to the sphere-sizing done in previous works~\cite{Daurer:2017} was used to determine the incident fluence using the scattering cross-section for gold at \SI{6}{\kilo\electronvolt}.

\subsubsection*{Phase retrieval}
Before performing phase retrieval, the intensity volumes were processed in the following manner: Background subtraction was performed using a rolling minimum filter with the window size 1.5 times the width of a fringe using the \texttt{ndimage.minimum\_filter} in \textit{SciPy}~\cite{Virtanen:2020}. Since no fringe contrast was visible at the outer resolution edges (see Supplementary Figure~\ref{fig:intensa}), the data was truncated such that the corners of the cube were within the sphere. The full-period resolution of the cube at the center-edge was \SI{1.61}{\nano\meter} and there were $384^3$ voxels for the three larger samples and $256^3$ voxels for the \texttt{cub17} sample. 

A combination of the error reduction (ER) algorithm and the difference map (DM) algorithm~\cite{Elser:2003} were used to reconstruct the electron densities from the background-subtracted intensity distribution. For each phasing run, 400 iterations were performed, 100 ER, 200 DM and 100 ER. Within each iteration, a dynamic real-space support constraint was applied by sorting the electron density values and only keeping the top $N_\mathrm{supp}$ values. The volume $N_\mathrm{supp}$ was chosen such that the histogram of densities inside the support had a small fraction of low values to ensure that the support was not too tight.

128 phasing runs from random white noise initial guesses were performed for each sample and the resulting densities were aligned and averaged to produce the final electron densities as well as the phase retrieval transfer function (PRTF). The PRTF strongly depends on the intensity at a voxel and thus exhibits an oscillatory behaviour along the fringes, which is not reflective of the quality of the structure at a given resolution since the lack of accurate phases near an interference minima barely affect the real-space structure. Thus, the 3D PRTF distribution was smoothed with a Gaussian kernel with a width half that of a fringe~\cite{Lundholm:2018}. The 3D isosurface plots were rendered using \textit{Chimera}~\cite{Pettersen:2004} and the contour levels were determined by using the ``Surface Color'' feature, colouring the surface by the gradient of the density and choosing the level which had the highest density gradients.

\subsection*{Acknowledgements}
We thank Rick Millane for helpful discussions. We acknowledge European XFEL in Schenefeld, Germany, for provision of X-ray free-electron laser beamtime at Scientific Instrument SPB/SFX (Single Particles, Clusters, and Biomolecules and Serial Femtosecond Crystallography) and would like to thank the staff for their assistance. We thank the DESY NanoLab, CSSB cryoEM user facility and the XBI labs at EuXFEL for access to electron microscopy resources and their staff for their help. This work has been supported by the Clusters of Excellence `Center for Ultrafast Imaging' (CUI,EXC 1074, ID 194651731) and `Advanced Imaging of Matter' (AIM,EXC 2056, ID 390715994) of the Deutsche Forschungsgemeinschaft (DFG). This work has also been supported by the European Research Council under the European Union's Seventh Framework Programme (FP7/2007-2013) through the Consolidator Grant COMOTION (ERC-614507-K{\"u}pper) and by the Helmholtz Gemeinschaft through the ``Impuls-und Vernetzungsfond''. J.C.H.S. and R.A.K. acknowledge support from the National Science Foundation BioXFEL award (STC-1231306). F.R.N.C.M. acknowledges support from the Swedish Research Council, R{\"o}ntgen-{\AA}ngstr{\"o}m Cluster and Carl Tryggers Foundation for Scientific Research. P.L.X. acknowledges a fellowship from the Joachim Herz Stiftung. P.L.X. and H.N.C. acknowledge support from the Human Frontiers Science Program (RGP0010/2017).

\subsection*{Author contributions}
The experiment was conceived by K.A. with the help of D.A.H., H.N.C, J.K{\"u}. and P.L.X.; P.L.X. prepared the samples with the help of H.L. and F.S.; J.B., K.G., R.B., H.K., R.L., T.S., M.S., P.V. and A.P.M. operated the SPB/SFX instrument at EuXFEL; J.B., A.K.S., R.A.K., M.S.H., A.D.E., J.L., L.W., N.R., S.A., D.A.H. and J.K{\"u}. setup and operated the sample delivery system; J.B., Y.K., R.L., T.S. and A.P.M. set up and operated the Rayleigh scattering diagnostics; B.J.D., F.R.N.C.M., and T.E. performed the online monitoring with help from K.A., T.W. and Z.S.; the data was analysed by K.A. with the help of Z.S., B.J.D., N.D.L., F.R.N.C.M., J.B., T.W., Y.Z., J.Ko., O.Y., A.B. and A.J.M.; K.A. wrote the manuscript with H.N.C., F.R.N.C.M., J.K{\"u} and J.H.C.S. with contributions from all authors.

\subsection*{Competing interests}
The authors declare no competing interests.

\printbibliography

@article{Li:2008,
  title={A facile polyol route to uniform gold octahedra with tailorable size and their optical properties},
  author={Li, Cuncheng and Shuford, Kevin L and Chen, Minghai and Lee, Eun Je and Cho, Sung Oh},
  journal={ACS nano},
  volume={2},
  number={9},
  pages={1760--1769},
  year={2008},
  publisher={ACS Publications}
}

@article{Lu:2017,
  title={Size-tunable uniform gold octahedra: fast synthesis, characterization, and plasmonic properties},
  author={Lu, Yonggang and Zhang, Haibin and Wu, Fan and Liu, Hong and Fang, Jingzhong},
  journal={RSC Advances},
  volume={7},
  number={30},
  pages={18601--18608},
  year={2017},
  publisher={Royal Society of Chemistry}
}

@article{Park:2018,
  title={Precisely Shaped, Uniformly Formed Gold Nanocubes with Ultrahigh Reproducibility in Single-Particle Scattering and Surface-Enhanced Raman Scattering},
  author={Park, Jeong-Eun and Lee, Yeonhee and Nam, Jwa-Min},
  journal={Nano letters},
  volume={18},
  number={10},
  pages={6475--6482},
  year={2018},
  publisher={ACS Publications}
}

@article{Reddy:2017,
author={Reddy, Hemanth K.N. and Yoon, Chun Hong and Aquila, Andrew and Awel, Salah and Ayyer, Kartik and Barty, Anton and Berntsen, Peter and Bielecki, Johan and Bobkov, Sergey and Bucher, Maximilian and Carini, Gabriella A. and Carron, Sebastian and Chapman, Henry and Daurer, Benedikt and DeMirci, Hasan and Ekeberg, Tomas and Fromme, Petra and Hajdu, Janos and Hanke, Max Felix and Hart, Philip and Hogue, Brenda G. and Hosseinizadeh, Ahmad and Kim, Yoonhee and Kirian, Richard A. and Kurta, Ruslan P. and Larsson, Daniel S.D. and Duane Loh, N. and Maia, Filipe R.N.C. and Mancuso, Adrian P. and M{\"u}hlig, Kerstin and Munke, Anna and Nam, Daewoong and Nettelblad, Carl and Ourmazd, Abbas and Rose, Max and Schwander, Peter and Seibert, Marvin and Sellberg, Jonas A. and Song, Changyong and Spence, John C.H. and Svenda, Martin and Van der Schot, Gijs and Vartanyants, Ivan A. and Williams, Garth J. and Xavier, P. Lourdu},
title={Coherent soft X-ray diffraction imaging of coliphage PR772 at the Linac coherent light source},
journal={Scientific Data},
year={2017},
%month={Jun},
day={27},
volume={4},
number={1},
pages={170079},
doi={10.1038/sdata.2017.79},
}

@article{Yoon:2011,
author = {Chun Hong Yoon and Peter Schwander and Chantal Abergel and Inger Andersson and Jakob Andreasson and Andrew Aquila and Sa\v{s}a Bajt and Miriam Barthelmess and Anton Barty and Michael J. Bogan and Christoph Bostedt and John Bozek and Henry N. Chapman and Jean-Michel Claverie and Nicola Coppola and Daniel P. DePonte and Tomas Ekeberg and Sascha W. Epp and Benjamin Erk and Holger Fleckenstein and Lutz Foucar and Heinz Graafsma and Lars Gumprecht and Janos Hajdu and Christina Y. Hampton and Andreas Hartmann and Elisabeth Hartmann and Robert Hartmann and Gunter Hauser and Helmut Hirsemann and Peter Holl and Stephan Kassemeyer and Nils Kimmel and Maya Kiskinova and Mengning Liang and Ne-Te Duane Loh and Lukas Lomb and Filipe R. N. C. Maia and Andrew V. Martin and Karol Nass and Emanuele Pedersoli and Christian Reich and Daniel Rolles and Benedikt Rudek and Artem Rudenko and Ilme Schlichting and Joachim Schulz and Marvin Seibert and Virginie Seltzer and Robert L. Shoeman and Raymond G. Sierra and Heike Soltau and Dmitri Starodub and Jan Steinbrener and Gunter Stier and Lothar Str\"{u}der and Martin Svenda and Joachim Ullrich and Georg Weidenspointner and Thomas A. White and Cornelia Wunderer and Abbas Ourmazd},
journal = {Opt. Express},
keywords = {X-ray imaging; Scattering, molecules; Ultrafast measurements; Edge detection; Far field diffraction; Free electron lasers; Silica; Spatial frequency; X ray diffraction},
number = {17},
pages = {16542--16549},
publisher = {OSA},
title = {Unsupervised classification of single-particle X-ray diffraction snapshots by spectral clustering},
volume = {19},
%month = {Aug},
year = {2011},
doi = {10.1364/OE.19.016542},
}

@article{Loh:2009,
  title={Reconstruction algorithm for single-particle diffraction imaging experiments},
  author={Loh, Ne-Te Duane and Elser, Veit},
  journal={Physical Review E},
  volume={80},
  number={2},
  pages={026705},
  year={2009},
  publisher={APS}
}

@article{Ayyer:2016,
  title={Dragonfly: an implementation of the expand--maximize--compress algorithm for single-particle imaging},
  author={Ayyer, Kartik and Lan, T-Y and Elser, Veit and Loh, N Duane},
  journal={Journal of applied crystallography},
  volume={49},
  number={4},
  pages={1320--1335},
  year={2016},
  publisher={International Union of Crystallography}
}

@article{Neutze:2000,
  title={Potential for biomolecular imaging with femtosecond X-ray pulses},
  author={Neutze, Richard and Wouts, Remco and Van der Spoel, David and Weckert, Edgar and Hajdu, Janos},
  journal={Nature},
  volume={406},
  number={6797},
  pages={752--757},
  year={2000},
  publisher={Nature Publishing Group}
}

@article{Elser:2003,
  title={Phase retrieval by iterated projections},
  author={Elser, Veit},
  journal={JOSA A},
  volume={20},
  number={1},
  pages={40--55},
  year={2003},
  publisher={Optical Society of America}
}

@article{Rose:2018,
  title={Single-particle imaging without symmetry constraints at an X-ray free-electron laser},
  author={Rose, Max and Bobkov, Sergey and Ayyer, Kartik and Kurta, Ruslan P. and Dzhigaev, Dmitry and Kim, Young Yong and Morgan, Andrew J. and Yoon, Chun Hong and Westphal, Daniel and Bielecki, Johan and Sellberg, Jonas A. and Williams, Garth and Maia, Filipe R.N.C. and Yefanov, Olexander M. and Ilyin, Vyacheslav and Mancuso, Adrian P. and Chapman, Henry N. and Hogue, Brenda G. and Aquila, Andrew and Barty, Anton and Vartanyants, Ivan A.},
  journal={IUCrJ},
  volume={5},
  number={6},
  pages={727--736},
  year={2018},
  publisher={International Union of Crystallography}
}

@article{Philipp:2012,
  title={Solving structure with sparse, randomly-oriented x-ray data},
  author={Philipp, Hugh T and Ayyer, Kartik and Tate, Mark W and Elser, Veit and Gruner, Sol M},
  journal={Optics express},
  volume={20},
  number={12},
  pages={13129--13137},
  year={2012},
  publisher={Optical Society of America}
}

@article{AyyerG:2015,
  title={Perspectives for imaging single protein molecules with the present design of the European XFEL},
  author={Ayyer, Kartik and Geloni, Gianluca and Kocharyan, Vitali and Saldin, Evgeni and Serkez, Svitozar and Yefanov, Oleksandr and Zagorodnov, Igor},
  journal={Structural dynamics},
  volume={2},
  number={4},
  pages={041702},
  year={2015},
  publisher={ACA}
}

@article{Ekeberg:2015,
  title = {Three-Dimensional Reconstruction of the Giant Mimivirus Particle with an X-Ray Free-Electron Laser},
  author = {Ekeberg, Tomas and Svenda, Martin and Abergel, Chantal and Maia, Filipe R. N. C. and Seltzer, Virginie and Claverie, Jean-Michel and Hantke, Max and J\"onsson, Olof and Nettelblad, Carl and van der Schot, Gijs and Liang, Mengning and DePonte, Daniel P. and Barty, Anton and Seibert, M. Marvin and Iwan, Bianca and Andersson, Inger and Loh, N. Duane and Martin, Andrew V. and Chapman, Henry and Bostedt, Christoph and Bozek, John D. and Ferguson, Ken R. and Krzywinski, Jacek and Epp, Sascha W. and Rolles, Daniel and Rudenko, Artem and Hartmann, Robert and Kimmel, Nils and Hajdu, Janos},
  journal = {Phys. Rev. Lett.},
  volume = {114},
  issue = {9},
  pages = {098102},
  numpages = {6},
  year = {2015},
  %month = {Mar},
  publisher = {American Physical Society},
  doi = {10.1103/PhysRevLett.114.098102},
}

@article{Loh:2010,
  title = {Cryptotomography: Reconstructing 3D Fourier Intensities from Randomly Oriented Single-Shot Diffraction Patterns},
  author = {Loh, N. D. and Bogan, M. J. and Elser, V. and Barty, A. and Boutet, S. and Bajt, S. and Hajdu, J. and Ekeberg, T. and Maia, F. R. N. C. and Schulz, J. and Seibert, M. M. and Iwan, B. and Timneanu, N. and Marchesini, S. and Schlichting, I. and Shoeman, R. L. and Lomb, L. and Frank, M. and Liang, M. and Chapman, H. N.},
  journal = {Phys. Rev. Lett.},
  volume = {104},
  issue = {22},
  pages = {225501},
  numpages = {5},
  year = {2010},
  %month = {Jun},
  publisher = {American Physical Society},
  doi = {10.1103/PhysRevLett.104.225501},
}

@article{Munke:2016,
  author={Munke, Anna and Andreasson, Jakob and Aquila, Andrew and Awel, Salah and Ayyer, Kartik and Barty, Anton and Bean, Richard J. and Berntsen, Peter and Bielecki, Johan and Boutet, S{\'e}bastien and Bucher, Maximilian and Chapman, Henry N. and Daurer, Benedikt J. and DeMirci, Hasan and Elser, Veit and Fromme, Petra and Hajdu, Janos and Hantke, Max F. and Higashiura, Akifumi and Hogue, Brenda G. and Hosseinizadeh, Ahmad and Kim, Yoonhee and Kirian, Richard A. and Reddy, Hemanth K. N. and Lan, Ti-Yen and Larsson, Daniel S. D. and Liu, Haiguang and Loh, N. Duane and Maia, Filipe R. N. C. and Mancuso, Adrian P. and M{\"u}hlig, Kerstin and Nakagawa, Atsushi and Nam, Daewoong and Nelson, Garrett and Nettelblad, Carl and Okamoto, Kenta and Ourmazd, Abbas and Rose, Max and van der Schot, Gijs and Schwander, Peter and Seibert, M. Marvin and Sellberg, Jonas A. and Sierra, Raymond G. and Song, Changyong and Svenda, Martin and Timneanu, Nicusor and Vartanyants, Ivan A. and Westphal, Daniel and Wiedorn, Max O. and Williams, Garth J. and Xavier, Paulraj Lourdu and Yoon, Chun Hong and Zook, James},
  title={Coherent diffraction of single Rice Dwarf virus particles using hard X-rays at the Linac Coherent Light Source},
  journal={Scientific Data},
  year={2016},
  %month={Aug},
  day={01},
  publisher={Nature Publishing Group},
  volume={3},
  pages={160064},
  note={Data Descriptor},
}

@article{Lundholm:2018,
author = "Lundholm, Ida V. and Sellberg, Jonas A. and Ekeberg, Tomas and Hantke, Max F. and Okamoto, Kenta and van der Schot, Gijs and Andreasson, Jakob and Barty, Anton and Bielecki, Johan and Bruza, Petr and Bucher, Max and Carron, Sebastian and Daurer, Benedikt J. and Ferguson, Ken and Hasse, Dirk and Krzywinski, Jacek and Larsson, Daniel S. D. and Morgan, Andrew and M{\"{u}}hlig, Kerstin and M{\"{u}}ller, Maria and Nettelblad, Carl and Pietrini, Alberto and Reddy, Hemanth K. N. and Rupp, Daniela and Sauppe, Mario and Seibert, Marvin and Svenda, Martin and Swiggers, Michelle and Timneanu, Nicusor and Ulmer, Anatoli and Westphal, Daniel and Williams, Garth and Zani, Alessandro and Faigel, Gyula and Chapman, Henry N. and M{\"{o}}ller, Thomas and Bostedt, Christoph and Hajdu, Janos and Gorkhover, Tais and Maia, Filipe R. N. C.",
title = "{Considerations for three-dimensional image reconstruction from experimental data in coherent diffractive imaging}",
journal = "IUCrJ",
year = "2018",
volume = "5",
number = "5",
pages = "531--541",
%month = "Sep",
doi = {10.1107/S2052252518010047},
keywords = {XFELs, Melbournevirus, coherent diffractive imaging, LCLS, image reconstruction},
}

@article{Ayyer:2019,
  title={Low-signal limit of X-ray single particle diffractive imaging},
  author={Ayyer, Kartik and Morgan, Andrew J and Aquila, Andrew and DeMirci, Hasan and Hogue, Brenda G and Kirian, Richard A and Xavier, P Lourdu and Yoon, Chun Hong and Chapman, Henry N and Barty, Anton},
  journal={Optics Express},
  volume={27},
  number={26},
  pages={37816--37833},
  year={2019},
  publisher={Optical Society of America}
}

@article{Bielecki:2019,
  author = {Bielecki, Johan and Hantke, Max F. and Daurer, Benedikt J. and Reddy, Hemanth K. N. and Hasse, Dirk and Larsson, Daniel S. D. and Gunn, Laura H. and Svenda, Martin and Munke, Anna and Sellberg, Jonas A. and Flueckiger, Leonie and Pietrini, Alberto and Nettelblad, Carl and Lundholm, Ida and Carlsson, Gunilla and Okamoto, Kenta and Timneanu, Nicusor and Westphal, Daniel and Kulyk, Olena and Higashiura, Akifumi and van der Schot, Gijs and Loh, Ne-Te Duane and Wysong, Taylor E. and Bostedt, Christoph and Gorkhover, Tais and Iwan, Bianca and Seibert, M. Marvin and Osipov, Timur and Walter, Peter and Hart, Philip and Bucher, Maximilian and Ulmer, Anatoli and Ray, Dipanwita and Carini, Gabriella and Ferguson, Ken R. and Andersson, Inger and Andreasson, Jakob and Hajdu, Janos and Maia, Filipe R. N. C.},
  title = {Electrospray sample injection for single-particle imaging with x-ray lasers},
  volume = {5},
  number = {5},
  elocation-id = {eaav8801},
  year = {2019},
  doi = {10.1126/sciadv.aav8801},
  publisher = {American Association for the Advancement of Science},
  journal = {Science Advances}
}

@article{Giewekemeyer:2019,
author = "Giewekemeyer, K. and Aquila, A. and Loh, N.-T. D. and Chushkin, Y. and Shanks, K. S. and Weiss, J.T. and Tate, M. W. and Philipp, H. T. and Stern, S. and Vagovic, P. and Mehrjoo, M. and Teo, C. and Barthelmess, M. and Zontone, F. and Chang, C. and Tiberio, R. C. and Sakdinawat, A. and Williams, G. J. and Gruner, S. M. and Mancuso, A. P.",
title = "{Experimental 3D coherent diffractive imaging from photon-sparse random projections}",
journal = "IUCrJ",
year = "2019",
volume = "6",
number = "3",
pages = "357--365",
%month = "May",
doi = {10.1107/S2052252519002781},
keywords = {coherent X-ray diffractive imaging (CXDI), X-ray free-electron lasers, XFELs, phase problem, single particles},
}

@article{Lan:2017,
author = "Lan, Ti-Yen and Wierman, Jennifer L. and Tate, Mark W. and Philipp, Hugh T. and Elser, Veit and Gruner, Sol M.",
title = "{Reconstructing three-dimensional protein crystal intensities from sparse unoriented two-axis X-ray diffraction patterns}",
journal = "Journal of Applied Crystallography",
year = "2017",
volume = "50",
number = "4",
pages = "985--993",
%month = "Aug",
doi = {10.1107/S1600576717006537},
abstract = {Recently, there has been a growing interest in adapting serial microcrystallography (SMX) experiments to existing storage ring (SR) sources. For very small crystals, however, radiation damage occurs before sufficient numbers of photons are diffracted to determine the orientation of the crystal. The challenge is to merge data from a large number of such `sparse' frames in order to measure the full reciprocal space intensity. To simulate sparse frames, a dataset was collected from a large lysozyme crystal illuminated by a dim X-ray source. The crystal was continuously rotated about two orthogonal axes to sample a subset of the rotation space. With the EMC algorithm [expand{--}maximize{--}compress; Loh & Elser (2009). {\it Phys. Rev. E}, {\bf 80}, 026705], it is shown that the diffracted intensity of the crystal can still be reconstructed even without knowledge of the orientation of the crystal in any sparse frame. Moreover, parallel computation implementations were designed to considerably improve the time and memory scaling of the algorithm. The results show that EMC-based SMX experiments should be feasible at SR sources.},
keywords = {X-ray serial microcrystallography, sparse data, EMC algorithm, protein microcrystallography, synchrotron radiation sources},
}

@article{Yefanov:2019,
author = {Yefanov, Oleksandr  and Oberthür, Dominik  and Bean, Richard  and Wiedorn, Max O.  and Knoska, Juraj  and Pena, Gisel  and Awel, Salah  and Gumprecht, Lars  and Domaracky, Martin  and Sarrou, Iosifina  and Lourdu Xavier, P.  and Metz, Markus  and Bajt, Saša  and Mariani, Valerio  and Gevorkov, Yaroslav  and White, Thomas A.  and Tolstikova, Aleksandra  and Villanueva-Perez, Pablo  and Seuring, Carolin  and Aplin, Steve  and Estillore, Armando D.  and Küpper, Jochen  and Klyuev, Alexander  and Kuhn, Manuela  and Laurus, Torsten  and Graafsma, Heinz  and Monteiro, Diana C. F.  and Trebbin, Martin  and Maia, Filipe R. N. C.  and Cruz-Mazo, Francisco  and Gañán-Calvo, Alfonso M.  and Heymann, Michael  and Darmanin, Connie  and Abbey, Brian  and Schmidt, Marius  and Fromme, Petra  and Giewekemeyer, Klaus  and Sikorski, Marcin  and Graceffa, Rita  and Vagovic, Patrik  and Kluyver, Thomas  and Bergemann, Martin  and Fangohr, Hans  and Sztuk-Dambietz, Jolanta  and Hauf, Steffen  and Raab, Natascha  and Bondar, Valerii  and Mancuso, Adrian P.  and Chapman, Henry  and Barty, Anton }, 
title = {Evaluation of serial crystallographic structure determination within megahertz pulse trains},
journal = {Structural Dynamics},
volume = {6},
number = {6},
pages = {064702},
year = {2019},
doi = {10.1063/1.5124387},
}

@article{Daurer:2017,
author = "Daurer, Benedikt J. and Okamoto, Kenta and Bielecki, Johan and Maia, Filipe R. N. C. and M{\"{u}}hlig, Kerstin and Seibert, M. Marvin and Hantke, Max F. and Nettelblad, Carl and Benner, W. Henry and Svenda, Martin and T{\^\i}mneanu, Nicu{\c{s}}or and Ekeberg, Tomas and Loh, N. Duane and Pietrini, Alberto and Zani, Alessandro and Rath, Asawari D. and Westphal, Daniel and Kirian, Richard A. and Awel, Salah and Wiedorn, Max O. and van der Schot, Gijs and Carlsson, Gunilla H. and Hasse, Dirk and Sellberg, Jonas A. and Barty, Anton and Andreasson, Jakob and Boutet, S{\'{e}}bastien and Williams, Garth and Koglin, Jason and Andersson, Inger and Hajdu, Janos and Larsson, Daniel S. D.",
title = "{Experimental strategies for imaging bioparticles with femtosecond hard X-ray pulses}",
journal = "IUCrJ",
year = "2017",
volume = "4",
number = "3",
pages = "251--262",
%month = "May",
doi = {10.1107/S2052252517003591},
}

@article{Roth:2018,
  title={Optimizing aerodynamic lenses for single-particle imaging},
  author={Roth, Nils and Awel, Salah and Horke, Daniel A and K{\"u}pper, Jochen},
  journal={Journal of Aerosol Science},
  volume={124},
  pages={17--29},
  year={2018},
  publisher={Elsevier}
}

@article{Hantke:2018,
  title={Rayleigh-scattering microscopy for tracking and sizing nanoparticles in focused aerosol beams},
  author={Hantke, Max F and Bielecki, Johan and Kulyk, Olena and Westphal, Daniel and Larsson, Daniel SD and Svenda, Martin and Reddy, Hemanth KN and Kirian, Richard A and Andreasson, Jakob and Hajdu, Janos and others},
  journal={IUCrJ},
  volume={5},
  number={6},
  pages={673--680},
  year={2018},
  publisher={International Union of Crystallography}
}

@article{Ueda:1998,
  title={Deterministic annealing EM algorithm},
  author={Ueda, Naonori and Nakano, Ryohei},
  journal={Neural networks},
  volume={11},
  number={2},
  pages={271--282},
  year={1998},
  publisher={Elsevier}
}

@article{Mancuso:2019,
  author = "Mancuso, Adrian P. and Aquila, Andrew and Batchelor, Lewis and Bean, Richard J. and Bielecki, Johan and Borchers, Gannon and Doerner, Katerina and Giewekemeyer, Klaus and Graceffa, Rita and Kelsey, Oliver D. and Kim, Yoonhee and Kirkwood, Henry J. and Legrand, Alexis and Letrun, Romain and Manning, Bradley and Lopez Morillo, Luis and Messerschmidt, Marc and Mills, Grant and Raabe, Steffen and Reimers, Nadja and Round, Adam and Sato, Tokushi and Schulz, Joachim and Signe Takem, Cedric and Sikorski, Marcin and Stern, Stephan and Thute, Prasad and Vagovi{\v{c}}, Patrik and Weinhausen, Britta and Tschentscher, Thomas",
  title = "{The Single Particles, Clusters and Biomolecules and Serial Femtosecond Crystallography instrument of the European XFEL: initial installation}",
  journal = "Journal of Synchrotron Radiation",
  year = "2019",
  volume = "26",
  number = "3",
  pages = "660--676",
  %month = "May",
  doi = {10.1107/S1600577519003308},
  keywords = {XFEL, serial crystallography, instrumentation},
}

@article{Heinrich:2011,
  title = "The adaptive gain integrating pixel detector AGIPD a detector for the European XFEL",
  journal = "Nuclear Instruments and Methods in Physics Research Section A: Accelerators, Spectrometers, Detectors and Associated Equipment",
  volume = "633",
  pages = "S11 - S14",
  year = "2011",
  note = "11th International Workshop on Radiation Imaging Detectors (IWORID)",
  doi = "https://doi.org/10.1016/j.nima.2010.06.107",
  author = "B. Henrich and J. Becker and R. Dinapoli and P. Goettlicher and H. Graafsma and H. Hirsemann and R. Klanner and H. Krueger and R. Mazzocco and A. Mozzanica and H. Perrey and G. Potdevin and B. Schmitt and X. Shi and A.K. Srivastava and U. Trunk and C. Youngman",
}

@Article{Decking:2020,
  author={Decking, W.
  and Abeghyan, S. and Abramian, P. and Abramsky, A. and Aguirre, A. and Albrecht, C. and Alou, P. and Altarelli, M. and Altmann, P. and Amyan, K. and Anashin, V.
  and Mommerz, M. and Monaco, L. and Montiel, C. and Moretti, M. and Morozov, I. and Morozov, P. and Mross, D. and others},
  title={A MHz-repetition-rate hard X-ray free-electron laser driven by a superconducting linear accelerator},
  journal={Nature Photonics},
  year={2020},
  %month={May},
  pages={391--397},
  day={18},
  doi={10.1038/s41566-020-0607-z},
}

@article{Daurer:2016,
  title={Hummingbird: monitoring and analyzing flash X-ray imaging experiments in real time},
  author={Daurer, Benedikt J and Hantke, Max F and Nettelblad, Carl and Maia, Filipe RNC},
  journal={Journal of applied crystallography},
  volume={49},
  number={3},
  pages={1042--1047},
  year={2016},
  publisher={International Union of Crystallography}
}

@inproceedings{Heisen:2013,
  title={Karabo: An integrated software framework combining control, data management, and scientific computing tasks},
  author={Heisen, Burkhard and Boukhelef, Djelloul and Esenov, Sergey and Hauf, Steffen and Kozlova, Iryna and Maia, Luis and Parenti, Andrea and Szuba, Janusz and Weger, Kerstin and Wrona, Krzysztof and Youngman, Christopher},
  booktitle={14th International Conference on Accelerator \& Large Experimental Physics Control Systems, ICALEPCS2013. San Francisco, CA},
  year={2013}
}

@article{Awel:2016,
  title={Visualizing aerosol-particle injection for diffractive-imaging experiments},
  author={Awel, Salah and Kirian, Richard A and Eckerskorn, Niko and Wiedorn, Max and Horke, Daniel A and Rode, Andrei V and K{\"u}pper, Jochen and Chapman, Henry N},
  journal={Optics express},
  volume={24},
  number={6},
  pages={6507--6521},
  year={2016},
  publisher={Optical Society of America}
}

@Article{Virtanen:2020,
author={Virtanen, Pauli and Gommers, Ralf and Oliphant, Travis E. and Haberland, Matt and Reddy, Tyler and Cournapeau, David and Burovski, Evgeni and Peterson, Pearu and Weckesser, Warren and Bright, Jonathan and van der Walt, St{\'e}fan J. and Brett, Matthew and Wilson, Joshua and Millman, K. Jarrod and Mayorov, Nikolay and Nelson, Andrew R. J. and Jones, Eric and Kern, Robert and Larson, Eric and Carey, C. J. and Polat, {\.{I}}lhan and Feng, Yu and Moore, Eric W. and VanderPlas, Jake and Laxalde, Denis and Perktold, Josef and Cimrman, Robert and Henriksen, Ian and Quintero, E. A. and Harris, Charles R. and Archibald, Anne M. and Ribeiro, Ant{\^o}nio H. and Pedregosa, Fabian and van Mulbregt, Paul and Vijaykumar, Aditya and Bardelli, Alessandro Pietro and Rothberg, Alex and Hilboll, Andreas and Kloeckner, Andreas and Scopatz, Anthony and Lee, Antony and Rokem, Ariel and Woods, C. Nathan and Fulton, Chad and Masson, Charles and H{\"a}ggstr{\"o}m, Christian and Fitzgerald, Clark and Nicholson, David A. and Hagen, David R. and Pasechnik, Dmitrii V. and Olivetti, Emanuele and Martin, Eric and Wieser, Eric and Silva, Fabrice and Lenders, Felix and Wilhelm, Florian and Young, G. and Price, Gavin A. and Ingold, Gert-Ludwig and Allen, Gregory E. and Lee, Gregory R. and Audren, Herv{\'e} and Probst, Irvin and Dietrich, J{\"o}rg P. and Silterra, Jacob and Webber, James T. and Slavi{\v{c}}, Janko and Nothman, Joel and Buchner, Johannes and Kulick, Johannes and Sch{\"o}nberger, Johannes L. and de Miranda Cardoso, Jos{\'e} Vin{\'i}cius and Reimer, Joscha and Harrington, Joseph and Rodr{\'i}guez, Juan Luis Cano and Nunez-Iglesias, Juan and Kuczynski, Justin and Tritz, Kevin and Thoma, Martin and Newville, Matthew and K{\"u}mmerer, Matthias and Bolingbroke, Maximilian and Tartre, Michael and Pak, Mikhail and Smith, Nathaniel J. and Nowaczyk, Nikolai and Shebanov, Nikolay and Pavlyk, Oleksandr and Brodtkorb, Per A. and Lee, Perry and McGibbon, Robert T. and Feldbauer, Roman and Lewis, Sam and Tygier, Sam and Sievert, Scott and Vigna, Sebastiano and Peterson, Stefan and More, Surhud and Pudlik, Tadeusz and Oshima, Takuya and Pingel, Thomas J. and Robitaille, Thomas P. and Spura, Thomas and Jones, Thouis R. and Cera, Tim and Leslie, Tim and Zito, Tiziano and Krauss, Tom and Upadhyay, Utkarsh and Halchenko, Yaroslav O. and V{\'a}zquez-Baeza, Yoshiki and 1.0 Contributors, SciPy},
title={SciPy 1.0: fundamental algorithms for scientific computing in Python},
journal={Nature Methods},
year={2020},
%month={Mar},
day={01},
volume={17},
number={3},
pages={261-272},
abstract={SciPy is an open-source scientific computing library for the Python programming language. Since its initial release in 2001, SciPy has become a de facto standard for leveraging scientific algorithms in Python, with over 600 unique code contributors, thousands of dependent packages, over 100,000 dependent repositories and millions of downloads per year. In this work, we provide an overview of the capabilities and development practices of SciPy 1.0 and highlight some recent technical developments.},
doi={10.1038/s41592-019-0686-2},
}

@article{Pettersen:2004,
  title={UCSF Chimera—a visualization system for exploratory research and analysis},
  author={Pettersen, Eric F and Goddard, Thomas D and Huang, Conrad C and Couch, Gregory S and Greenblatt, Daniel M and Meng, Elaine C and Ferrin, Thomas E},
  journal={Journal of computational chemistry},
  volume={25},
  number={13},
  pages={1605--1612},
  year={2004},
  publisher={Wiley Online Library}
}

@Article{Barty:2012,
author={Barty, Anton and Caleman, Carl and Aquila, Andrew and Timneanu, Nicusor and Lomb, Lukas and White, Thomas A. and Andreasson, Jakob and Arnlund, David and Bajt, Sa{\v{s}}a and Barends, Thomas R. M. and Barthelmess, Miriam and Bogan, Michael J. and Bostedt, Christoph and Bozek, John D. and Coffee, Ryan and Coppola, Nicola and Davidsson, Jan and DePonte, Daniel P. and Doak, R. Bruce and Ekeberg, Tomas and Elser, Veit and Epp, Sascha W. and Erk, Benjamin and Fleckenstein, Holger and Foucar, Lutz and Fromme, Petra and Graafsma, Heinz and Gumprecht, Lars and Hajdu, Janos and Hampton, Christina Y. and Hartmann, Robert and Hartmann, Andreas and Hauser, G{\"u}nter and Hirsemann, Helmut and Holl, Peter and Hunter, Mark S. and Johansson, Linda and Kassemeyer, Stephan and Kimmel, Nils and Kirian, Richard A. and Liang, Mengning and Maia, Filipe R. N. C. and Malmerberg, Erik and Marchesini, Stefano and Martin, Andrew V. and Nass, Karol and Neutze, Richard and Reich, Christian and Rolles, Daniel and Rudek, Benedikt and Rudenko, Artem and Scott, Howard and Schlichting, Ilme and Schulz, Joachim and Seibert, M. Marvin and Shoeman, Robert L. and Sierra, Raymond G. and Soltau, Heike and Spence, John C. H. and Stellato, Francesco and Stern, Stephan and Str{\"u}der, Lothar and Ullrich, Joachim and Wang, X. and Weidenspointner, Georg and Weierstall, Uwe and Wunderer, Cornelia B. and Chapman, Henry N.},
title={Self-terminating diffraction gates femtosecond X-ray nanocrystallography measurements},
journal={Nature Photonics},
year={2012},
%month={Jan},
day={01},
volume={6},
number={1},
pages={35-40},
doi={10.1038/nphoton.2011.297},
}

@Article{Chapman:2006,
author={Chapman, Henry N. and Barty, Anton and Bogan, Michael J. and Boutet, S{\'e}bastien and Frank, Matthias and Hau-Riege, Stefan P. and Marchesini, Stefano and Woods, Bruce W. and Bajt, Sa{\v{s}}a and Benner, W. Henry and London, Richard A. and Pl{\"o}njes, Elke and Kuhlmann, Marion and Treusch, Rolf and D{\"u}sterer, Stefan and Tschentscher, Thomas and Schneider, Jochen R. and Spiller, Eberhard and M{\"o}ller, Thomas and Bostedt, Christoph and Hoener, Matthias and Shapiro, David A. and Hodgson, Keith O. and van der Spoel, David and Burmeister, Florian and Bergh, Magnus and Caleman, Carl and Huldt, G{\"o}sta and Seibert, M. Marvin and Maia, Filipe R. N. C. and Lee, Richard W. and Sz{\"o}ke, Abraham and Timneanu, Nicusor and Hajdu, Janos},
title={Femtosecond diffractive imaging with a soft-X-ray free-electron laser},
journal={Nature Physics},
year={2006},
%month={Dec},
day={01},
volume={2},
number={12},
pages={839-843},
abstract={Theory predicts1,2,3,4 that, with an ultrashort and extremely bright coherent X-ray pulse, a single diffraction pattern may be recorded from a large macromolecule, a virus or a cell before the sample explodes and turns into a plasma. Here we report the first experimental demonstration of this principle using the FLASH soft-X-ray free-electron laser. An intense 25{\thinspace}fs, 4{\texttimes}1013{\thinspace}W{\thinspace}cm−2 pulse, containing 1012 photons at 32{\thinspace}nm wavelength, produced a coherent diffraction pattern from a nanostructured non-periodic object, before destroying it at 60,000{\thinspace}K. A novel X-ray camera assured single-photon detection sensitivity by filtering out parasitic scattering and plasma radiation. The reconstructed image, obtained directly from the coherent pattern by phase retrieval through oversampling5,6,7,8,9, shows no measurable damage, and is reconstructed at the diffraction-limited resolution. A three-dimensional data set may be assembled from such images when copies of a reproducible sample are exposed to the beam one by one10.},
doi={10.1038/nphys461},
}

@article{Fortmann-Grote:2017,
author = "Fortmann-Grote, Carsten and Buzmakov, Alexey and Jurek, Zoltan and Loh, Ne-Te Duane and Samoylova, Liubov and Santra, Robin and Schneidmiller, Evgeny A. and Tschentscher, Thomas and Yakubov, Sergey and Yoon, Chun Hong and Yurkov, Michael V. and Ziaja-Motyka, Beata and Mancuso, Adrian P.",
title = "{Start-to-end simulation of single-particle imaging using ultra-short pulses at the European X-ray Free-Electron Laser}",
journal = "IUCrJ",
year = "2017",
volume = "4",
number = "5",
pages = "560--568",
%month = "Sep",
doi = {10.1107/S2052252517009496},
keywords = {single-particle imaging, X-ray free-electron lasers, simulations, diffraction, scattering},
}

@article{Ayyer:2020,
  title={Reference-enhanced x-ray single-particle imaging},
  author={Ayyer, Kartik},
  journal={Optica},
  volume={7},
  number={6},
  pages={593--601},
  year={2020},
  publisher={Optical Society of America}
}

@article{Sobolev:2020,
author={Sobolev, Egor and Zolotarev, Sergei and Giewekemeyer, Klaus and Bielecki, Johan and Okamoto, Kenta and Reddy, Hemanth K. N. and Andreasson, Jakob and Ayyer, Kartik and Barak, Imrich and Bari, Sadia and Barty, Anton and Bean, Richard and Bobkov, Sergey and Chapman, Henry N. and Chojnowski, Grzegorz and Daurer, Benedikt J. and D{\"o}rner, Katerina and Ekeberg, Tomas and Fl{\"u}ckiger, Leonie and Galzitskaya, Oxana and Gelisio, Luca and Hauf, Steffen and Hogue, Brenda G. and Horke, Daniel A. and Hosseinizadeh, Ahmad and Ilyin, Vyacheslav and Jung, Chulho and Kim, Chan and Kim, Yoonhee and Kirian, Richard A. and Kirkwood, Henry and Kulyk, Olena and K{\"u}pper, Jochen and Letrun, Romain and Loh, N. Duane and Lorenzen, Kristina and Messerschmidt, Marc and M{\"u}hlig, Kerstin and Ourmazd, Abbas and Raab, Natascha and Rode, Andrei V. and Rose, Max and Round, Adam and Sato, Takushi and Schubert, Robin and Schwander, Peter and Sellberg, Jonas A. and Sikorski, Marcin and Silenzi, Alessandro and Song, Changyong and Spence, John C. H. and Stern, Stephan and Sztuk-Dambietz, Jolanta and Teslyuk, Anthon and Timneanu, Nicusor and Trebbin, Martin and Uetrecht, Charlotte and Weinhausen, Britta and Williams, Garth J. and Xavier, P. Lourdu and Xu, Chen and Vartanyants, Ivan A. and Lamzin, Victor S. and Mancuso, Adrian and Maia, Filipe R. N. C.},
title={Megahertz single-particle imaging at the European XFEL},
journal={Communications Physics},
year={2020},
%month={May},
day={29},
volume={3},
number={1},
pages={97},
doi={10.1038/s42005-020-0362-y},
}

@article{Jonsson:2015,
author = "J{\"{o}}nsson, H. Olof and Timneanu, Nicu{\c{s}}or and {\"{O}}stlin, Christofer and Scott, Howard A. and Caleman, Carl",
title = "{Simulations of radiation damage as a function of the~temporal pulse profile in femtosecond X-ray protein crystallography}",
journal = "Journal of Synchrotron Radiation",
year = "2015",
volume = "22",
number = "2",
pages = "256--266",
%month = "Mar",
doi = {10.1107/S1600577515002878},
keywords = {X-ray free-electron laser, serial femtosecond crystallography, radiation damage, plasma simulations},
}

@article{Scheres:2005,
  title={Maximum-likelihood multi-reference refinement for electron microscopy images},
  author={Scheres, Sjors HW and Valle, Mikel and Nu{\~n}ez, Rafael and Sorzano, Carlos OS and Marabini, Roberto and Herman, Gabor T and Carazo, Jose-Maria},
  journal={Journal of molecular biology},
  volume={348},
  number={1},
  pages={139--149},
  year={2005},
  publisher={Elsevier}
}

@Article{Ho:2020,
author={Ho, Phay J. and Daurer, Benedikt J. and Hantke, Max F. and Bielecki, Johan and Al Haddad, Andre and Bucher, Maximilian and Doumy, Gilles and Ferguson, Ken R. and Fl{\"u}ckiger, Leonie and Gorkhover, Tais and Iwan, Bianca and Knight, Christopher and Moeller, Stefan and Osipov, Timur and Ray, Dipanwita and Southworth, Stephen H. and Svenda, Martin and Timneanu, Nicusor and Ulmer, Anatoli and Walter, Peter and Hajdu, Janos and Young, Linda and Maia, Filipe R. N. C. and Bostedt, Christoph},
title={The role of transient resonances for ultra-fast imaging of single sucrose nanoclusters},
journal={Nature Communications},
year={2020},
%month={Jan},
day={09},
volume={11},
number={1},
pages={167},
doi={10.1038/s41467-019-13905-9},
}

@article{vanderwalt:2014,
  title={scikit-image: image processing in Python},
  author={Van der Walt, Stefan and Sch{\"o}nberger, Johannes L and Nunez-Iglesias, Juan and Boulogne, Fran{\c{c}}ois and Warner, Joshua D and Yager, Neil and Gouillart, Emmanuelle and Yu, Tony and the scikit-image contributors},
  journal={PeerJ},
  volume={2},
  pages={e453},
  year={2014},
  publisher={PeerJ Inc.}
}

@article{Henderson:1995,
  title={The potential and limitations of neutrons, electrons and X-rays for atomic resolution microscopy of unstained biological molecules},
  author={Henderson, Richard},
  journal={Quarterly reviews of biophysics},
  volume={28},
  number={2},
  pages={171--193},
  year={1995},
  publisher={Cambridge University Press}
}

@Article{Seibert:2011,
author={Seibert, M. Marvin and Ekeberg, Tomas and Maia, Filipe R. N. C. and Svenda, Martin and Andreasson, Jakob and J{\"o}nsson, Olof and Odi{\'{c}}, Du{\v{s}}ko and Iwan, Bianca and Rocker, Andrea and Westphal, Daniel and Hantke, Max and DePonte, Daniel P. and Barty, Anton and Schulz, Joachim and Gumprecht, Lars and Coppola, Nicola and Aquila, Andrew and Liang, Mengning and White, Thomas A. and Martin, Andrew and Caleman, Carl and Stern, Stephan and Abergel, Chantal and Seltzer, Virginie and Claverie, Jean-Michel and Bostedt, Christoph and Bozek, John D. and Boutet, S{\'e}bastien and Miahnahri, A. Alan and Messerschmidt, Marc and Krzywinski, Jacek and Williams, Garth and Hodgson, Keith O. and Bogan, Michael J. and Hampton, Christina Y. and Sierra, Raymond G. and Starodub, Dmitri and Andersson, Inger and Bajt, Sa{\v{s}}a and Barthelmess, Miriam and Spence, John C. H. and Fromme, Petra and Weierstall, Uwe and Kirian, Richard and Hunter, Mark and Doak, R. Bruce and Marchesini, Stefano and Hau-Riege, Stefan P. and Frank, Matthias and Shoeman, Robert L. and Lomb, Lukas and Epp, Sascha W. and Hartmann, Robert and Rolles, Daniel and Rudenko, Artem and Schmidt, Carlo and Foucar, Lutz and Kimmel, Nils and Holl, Peter and Rudek, Benedikt and Erk, Benjamin and H{\"o}mke, Andr{\'e} and Reich, Christian and Pietschner, Daniel and Weidenspointner, Georg and Str{\"u}der, Lothar and Hauser, G{\"u}nter and Gorke, Hubert and Ullrich, Joachim and Schlichting, Ilme and Herrmann, Sven and Schaller, Gerhard and Schopper, Florian and Soltau, Heike and K{\"u}hnel, Kai-Uwe and Andritschke, Robert and Schr{\"o}ter, Claus-Dieter and Krasniqi, Faton and Bott, Mario and Schorb, Sebastian and Rupp, Daniela and Adolph, Marcus and Gorkhover, Tais and Hirsemann, Helmut and Potdevin, Guillaume and Graafsma, Heinz and Nilsson, Bj{\"o}rn and Chapman, Henry N. and Hajdu, Janos},
title={Single mimivirus particles intercepted and imaged with an X-ray laser},
journal={Nature},
year={2011},
month={Feb},
day={01},
volume={470},
number={7332},
pages={78-81},
doi={10.1038/nature09748},
}

@article{DeCarlo:2004,
  title={Particle morphology and density characterization by combined mobility and aerodynamic diameter measurements. Part 1: Theory},
  author={DeCarlo, Peter F and Slowik, Jay G and Worsnop, Douglas R and Davidovits, Paul and Jimenez, Jose L},
  journal={Aerosol Science and Technology},
  volume={38},
  number={12},
  pages={1185--1205},
  year={2004},
  publisher={Taylor \& Francis}
}

@article{Bogan:2008,
author = {Bogan, Michael J. and Benner, W. Henry and Boutet, Sébastien and Rohner, Urs and Frank, Matthias and Barty, Anton and Seibert, M. Marvin and Maia, Filipe and Marchesini, Stefano and Bajt, Saša and Woods, Bruce and Riot, Vincent and Hau-Riege, Stefan P. and Svenda, Martin and Marklund, Erik and Spiller, Eberhard and Hajdu, Janos and Chapman, Henry N.},
title = {Single Particle X-ray Diffractive Imaging},
journal = {Nano Letters},
volume = {8},
number = {1},
pages = {310-316},
year = {2008},
doi = {10.1021/nl072728k},
}

@article{Samanta:2020,
  title={Controlled beams of shock-frozen, isolated, biological and artificial nanoparticles},
  author={Samanta, Amit K and Amin, Muhamed and Estillore, Armando D and Roth, Nils and Worbs, Lena and Horke, Daniel A and K{\"u}pper, Jochen},
  journal={Structural Dynamics},
  volume={7},
  number={2},
  pages={024304},
  year={2020},
  publisher={American Crystallographic Association}
}

@article{Chang:2015,
  title={Spatially-controlled complex molecules and their applications},
  author={Chang, Yuan-Pin and Horke, Daniel A and Trippel, Sebastian and K{\"u}pper, Jochen},
  journal={International reviews in physical chemistry},
  volume={34},
  number={4},
  pages={557--590},
  year={2015},
  publisher={Taylor \& Francis}
}

\newpage

\renewcommand{\figurename}{Supplementary Figure}
\setcounter{figure}{0}
\section*{Supplementary Information}

\begin{figure}
  \centering
  \begin{tabular}{c c}
    \includegraphics[width=0.45\textwidth]{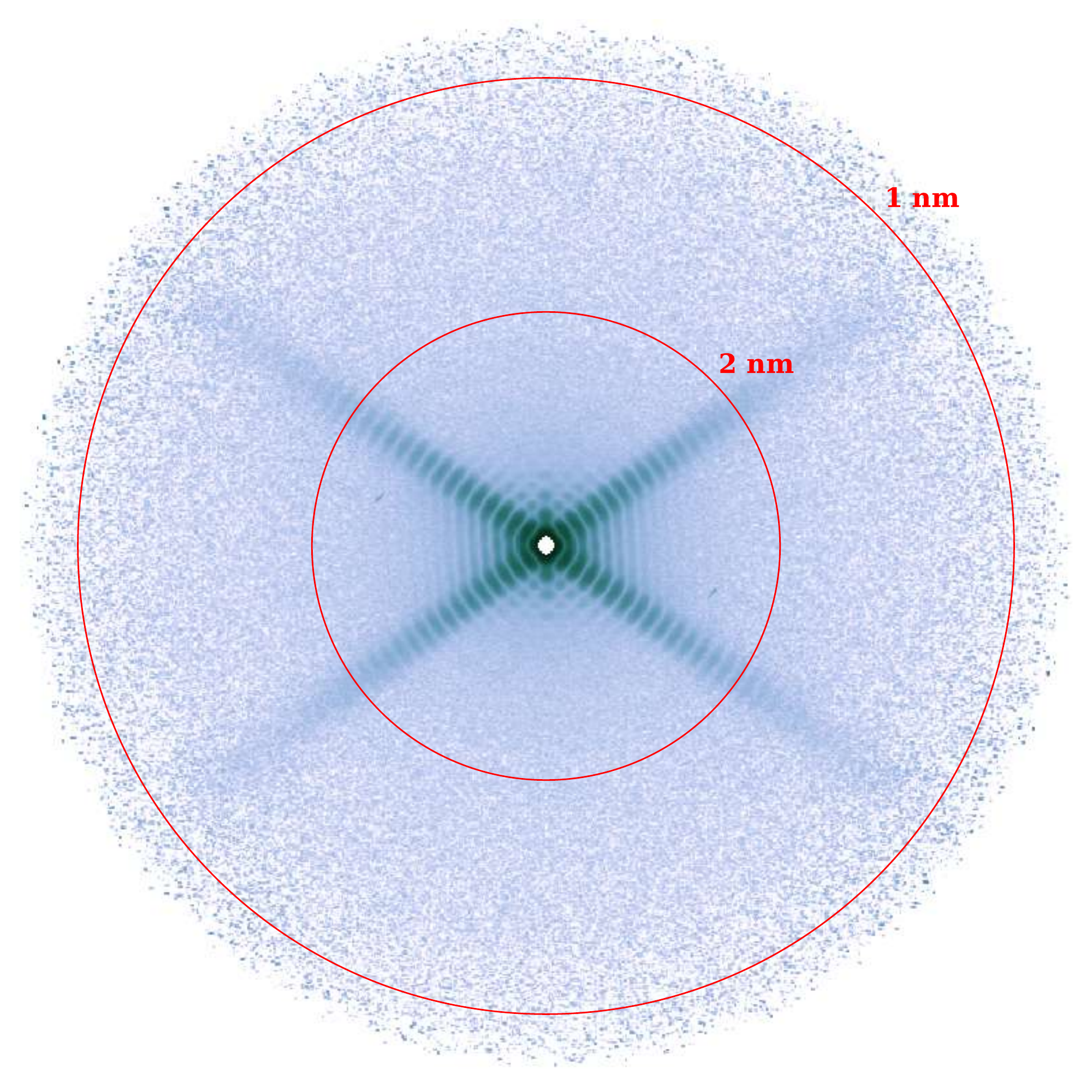} &
    \includegraphics[width=0.45\textwidth]{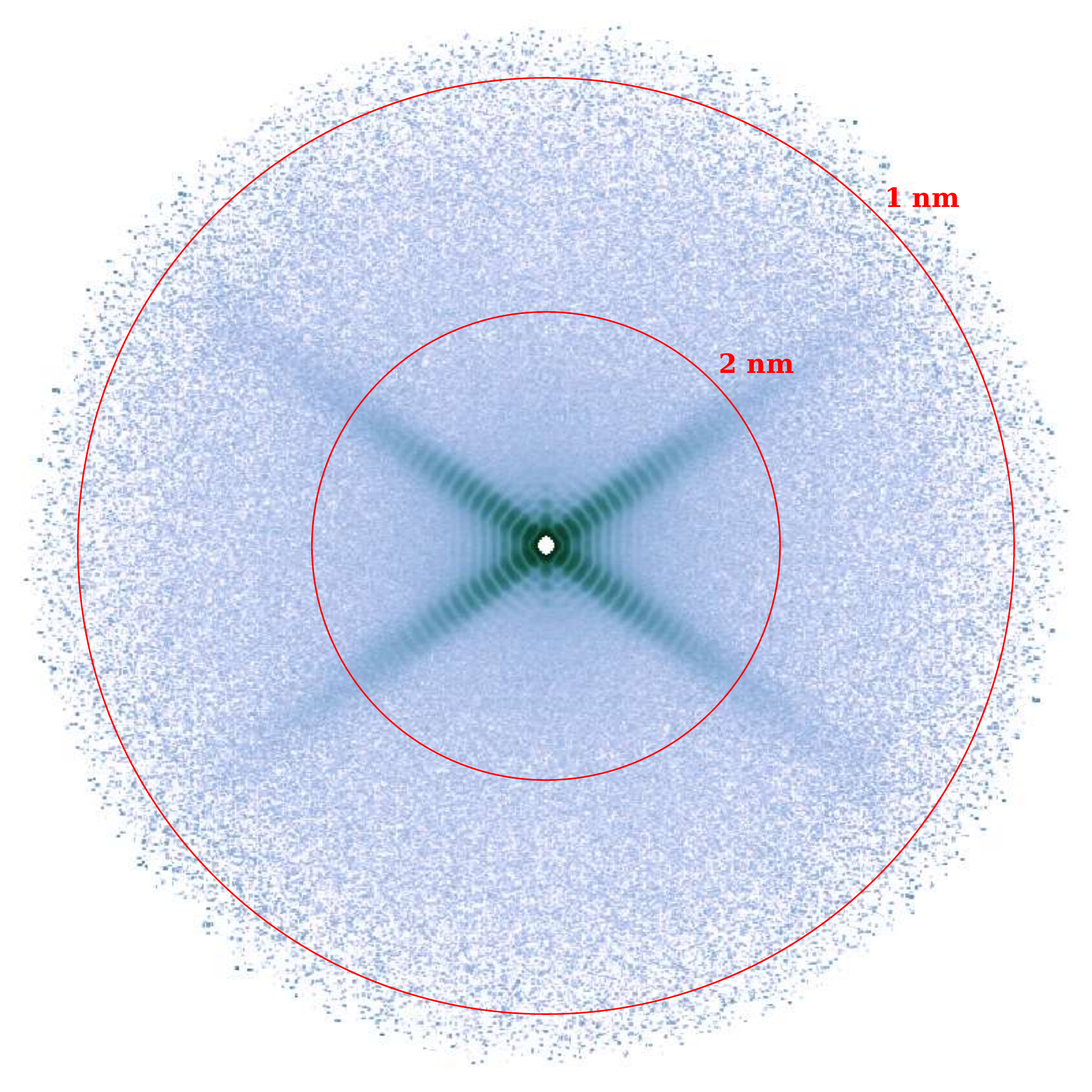} \\
    (a) & (b) \\
    \includegraphics[width=0.45\textwidth]{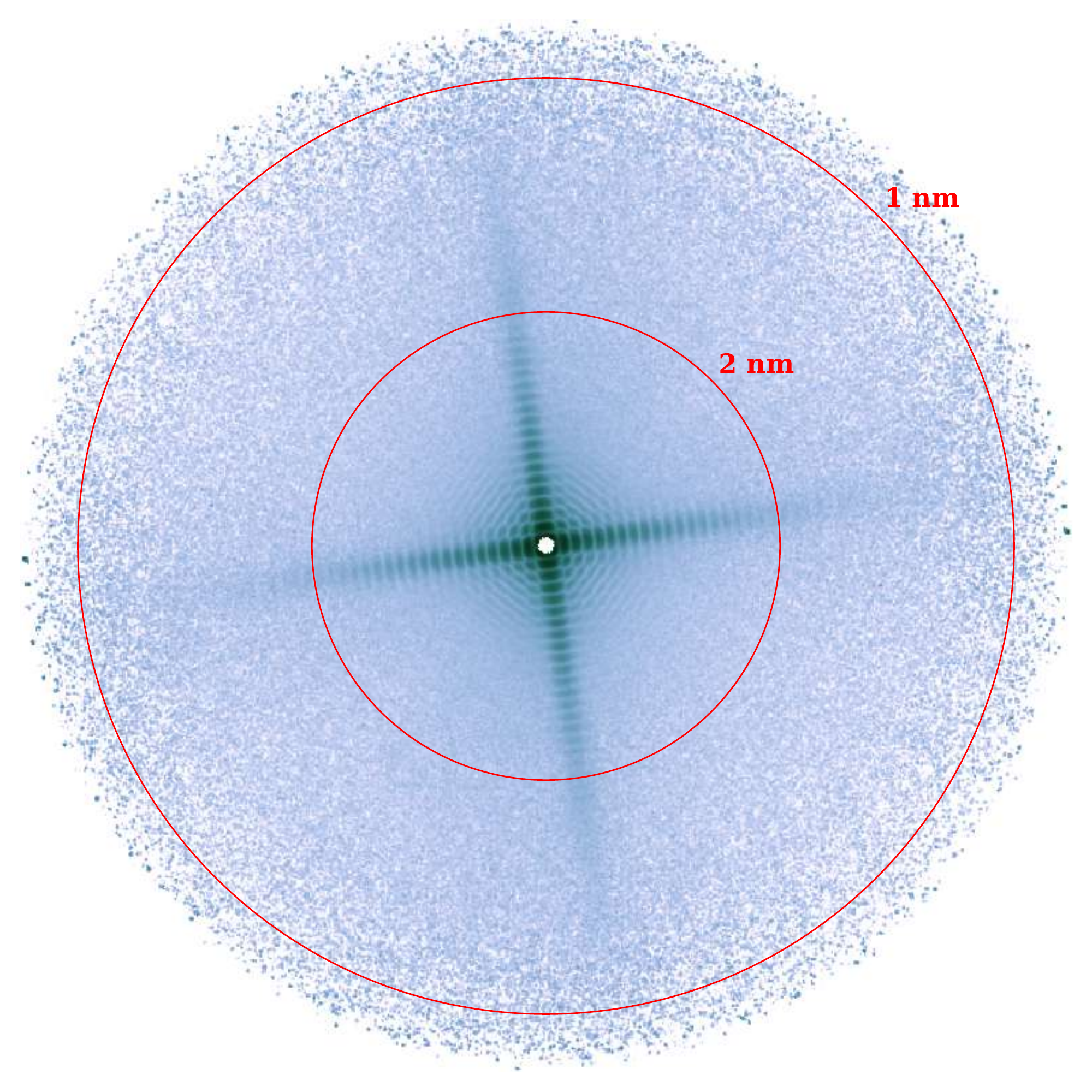} &
    \includegraphics[width=0.45\textwidth]{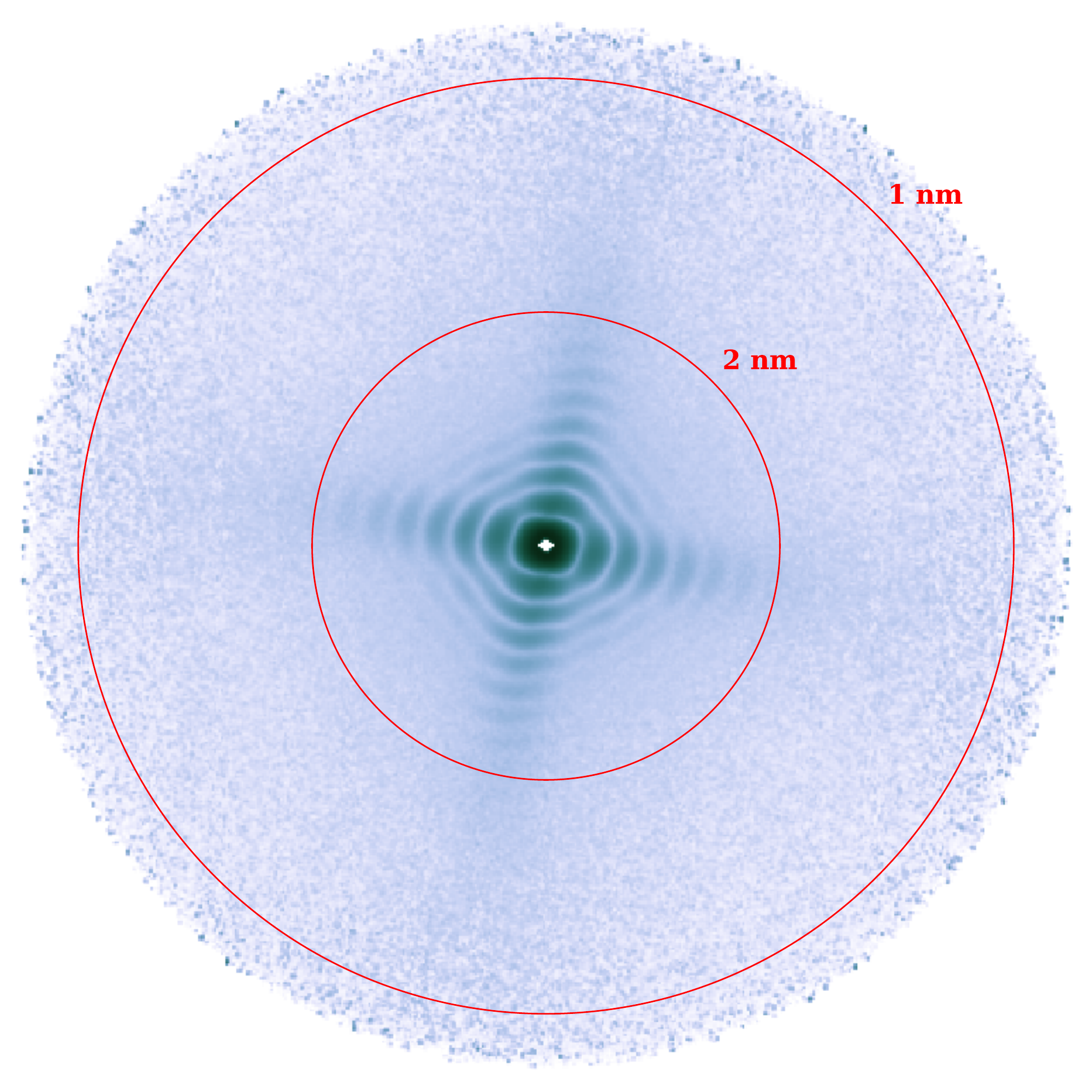} \\
    (c) & (d) \\
  \end{tabular}
  \caption{Full resolution intensity slices. Logarithmic intensity slices through intensity reconstructions using the full detector for the (a) \texttt{oct30} (b) \texttt{oct40} (c) \texttt{cub42} and (d) \texttt{cub17} datasets. The octahedral intensities were generated after structural sorting. The rings indicate the full-period resolution.}
  \label{fig:intensa}
\end{figure}

\begin{figure}
  \centering
  \begin{tabular}{c c}
    \includegraphics[width=0.45\textwidth]{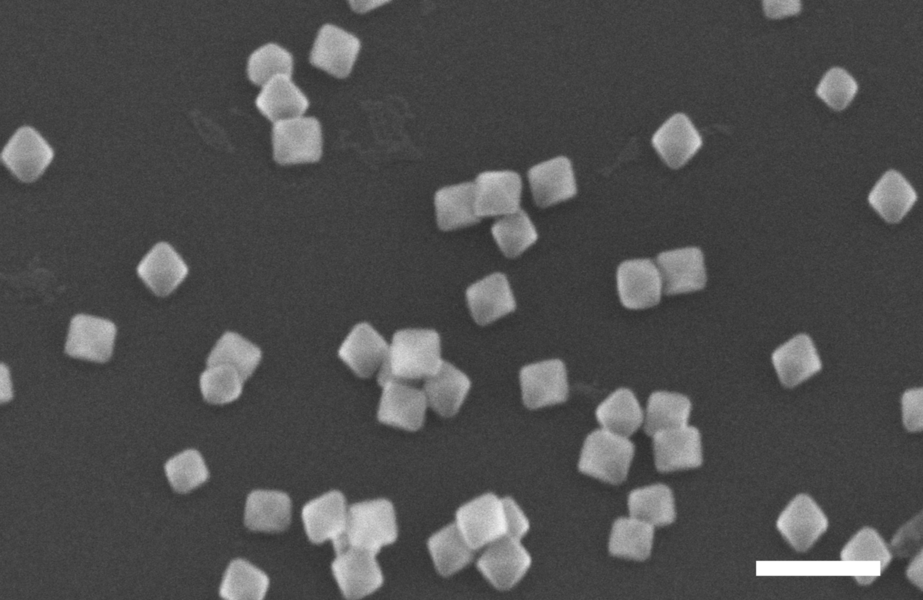} &
    \includegraphics[width=0.45\textwidth]{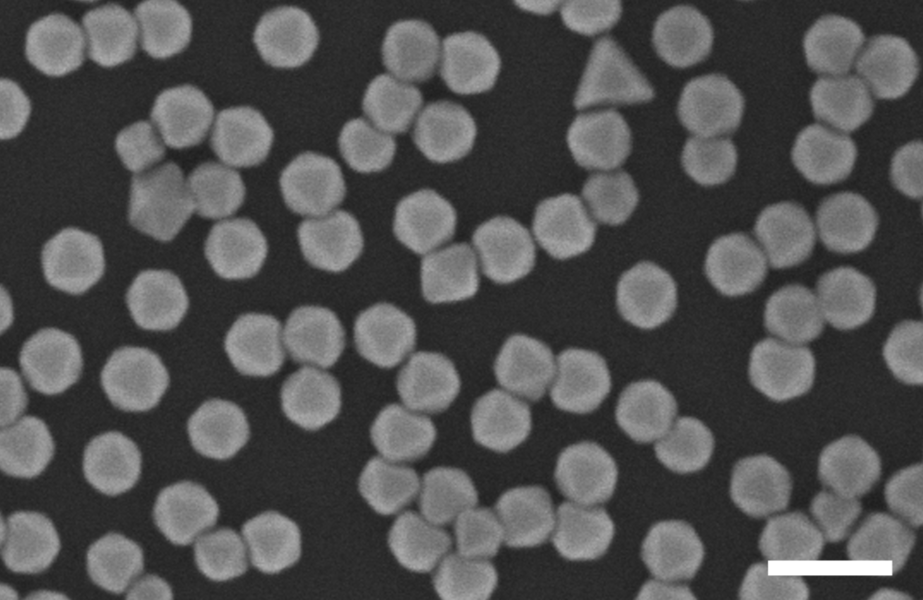} \\
    (a) & (b) \\
    \includegraphics[width=0.45\textwidth]{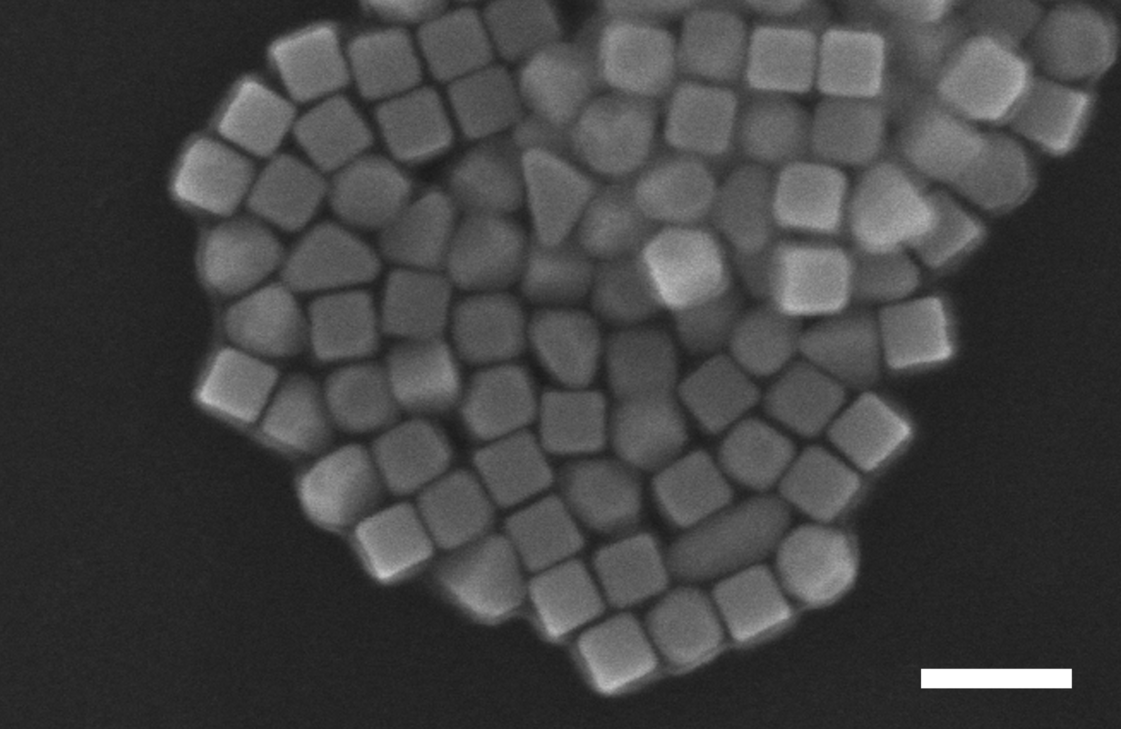} &
    \includegraphics[width=0.45\textwidth]{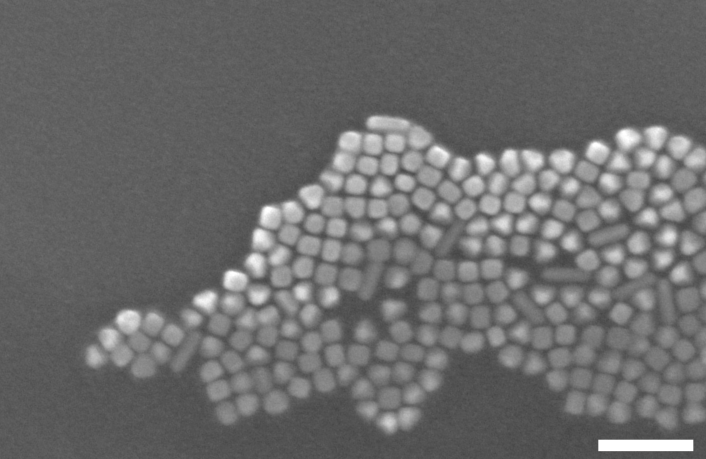} \\
    (c) & (d) \\
  \end{tabular}
  \caption{Scanning electron microscopy images. (a) \texttt{oct30} (b) \texttt{oct40} (c) \texttt{cub42} (d) \texttt{cub17}. All scale bars are \SI{100}{\nano\meter}.}
  \label{fig:sem_sample}
\end{figure}

\begin{figure}
  \centering
  \begin{tabular}{c c}
    \includegraphics[width=0.45\textwidth]{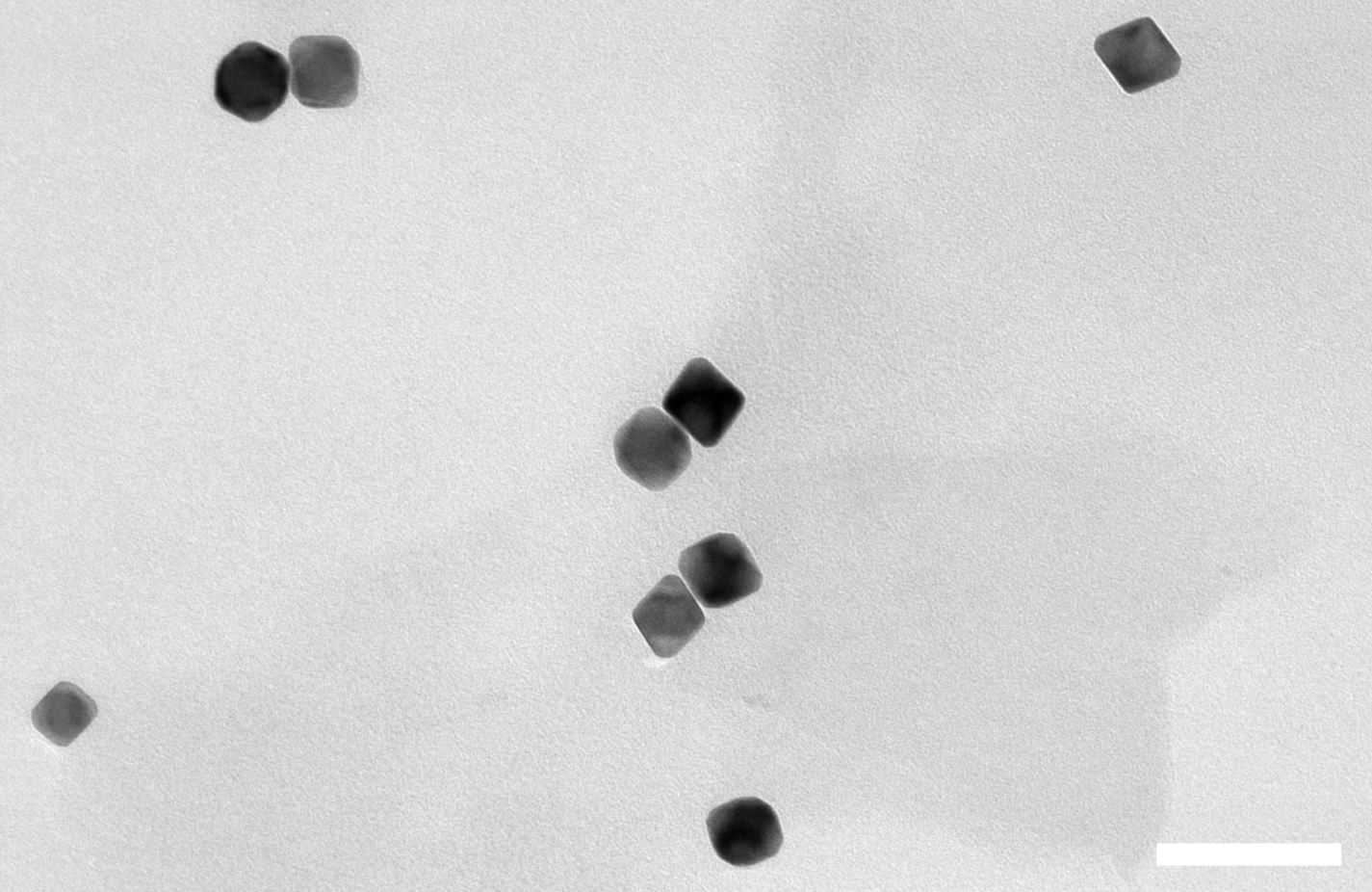} &
    \includegraphics[width=0.45\textwidth]{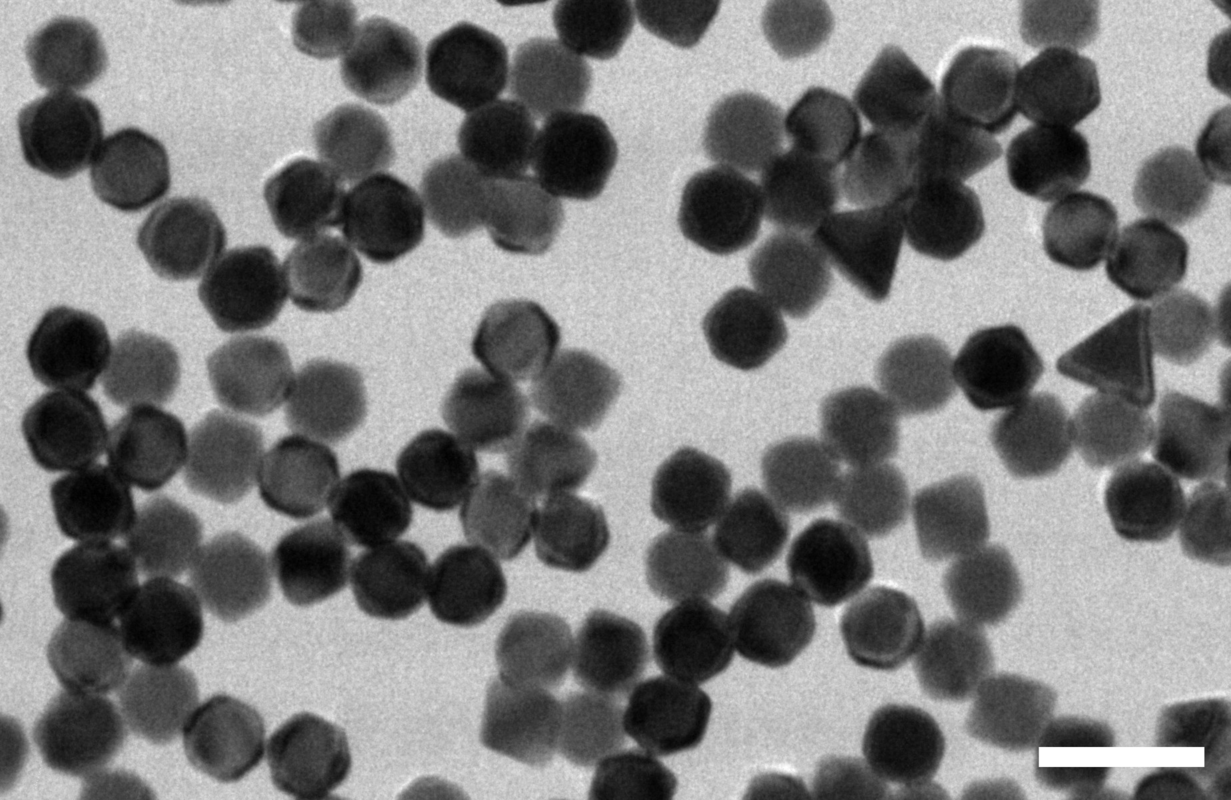} \\
    (a) & (b) \\
    \includegraphics[width=0.45\textwidth]{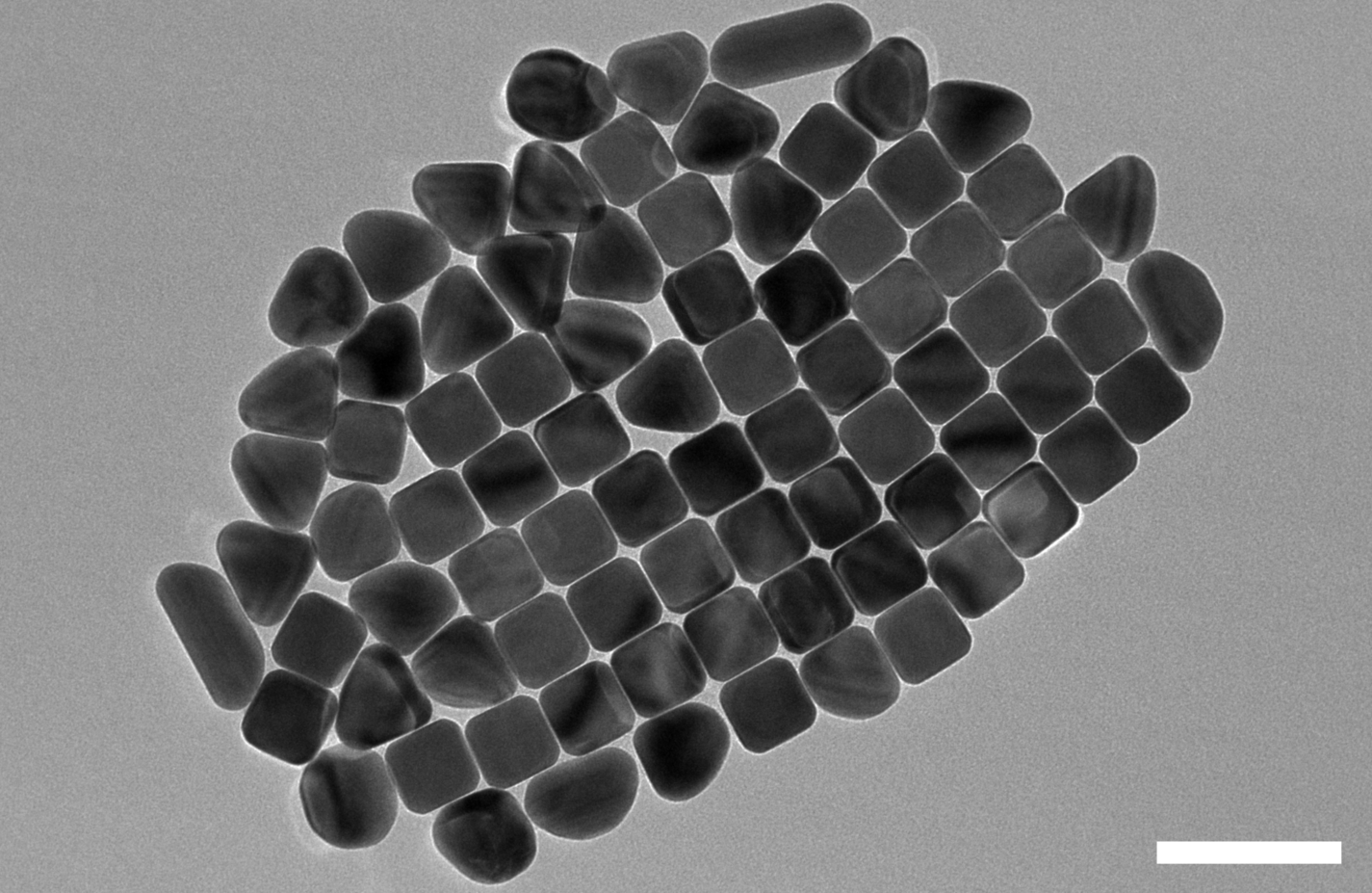} &
    \includegraphics[width=0.45\textwidth]{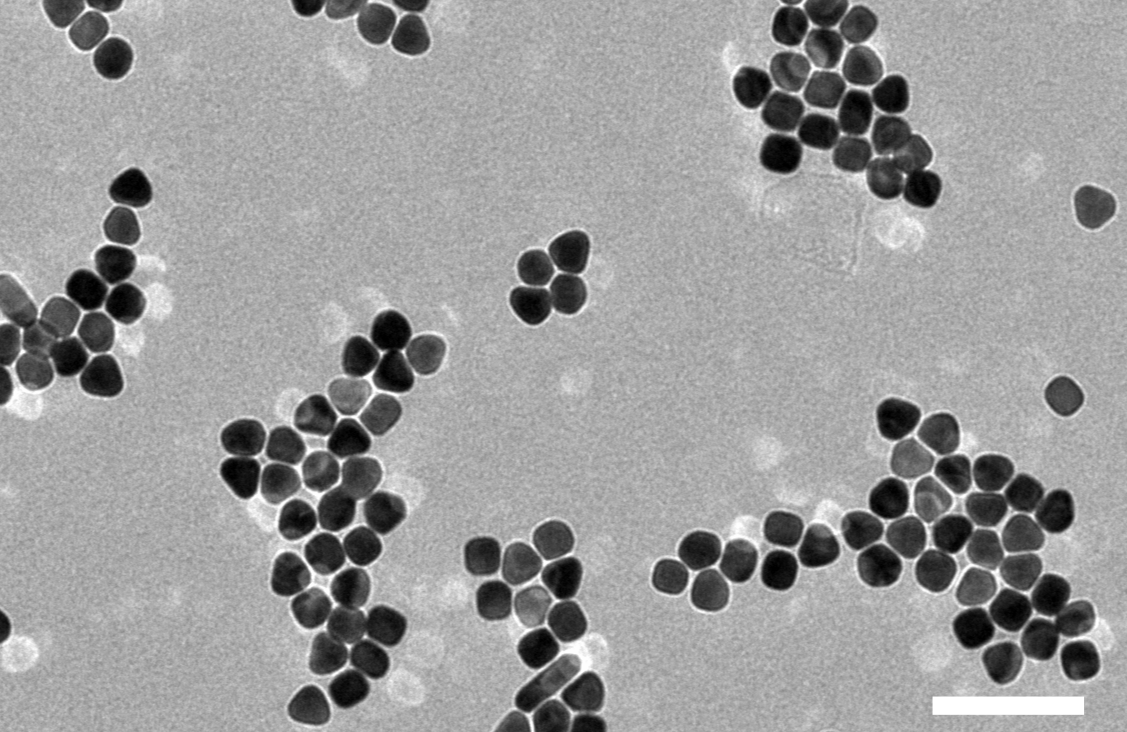} \\
    (c) & (d) \\
  \end{tabular}
  \caption{Transmission electron microscopy images. (a) \texttt{oct30} (b) \texttt{oct40} (c) \texttt{cub42} (d) \texttt{cub17}. All scale bars are \SI{100}{\nano\meter}.}
  \label{fig:tem_sample}
\end{figure}

\begin{figure}
  \centering
  \begin{tabular}{c c}
    \includegraphics[width=0.45\textwidth]{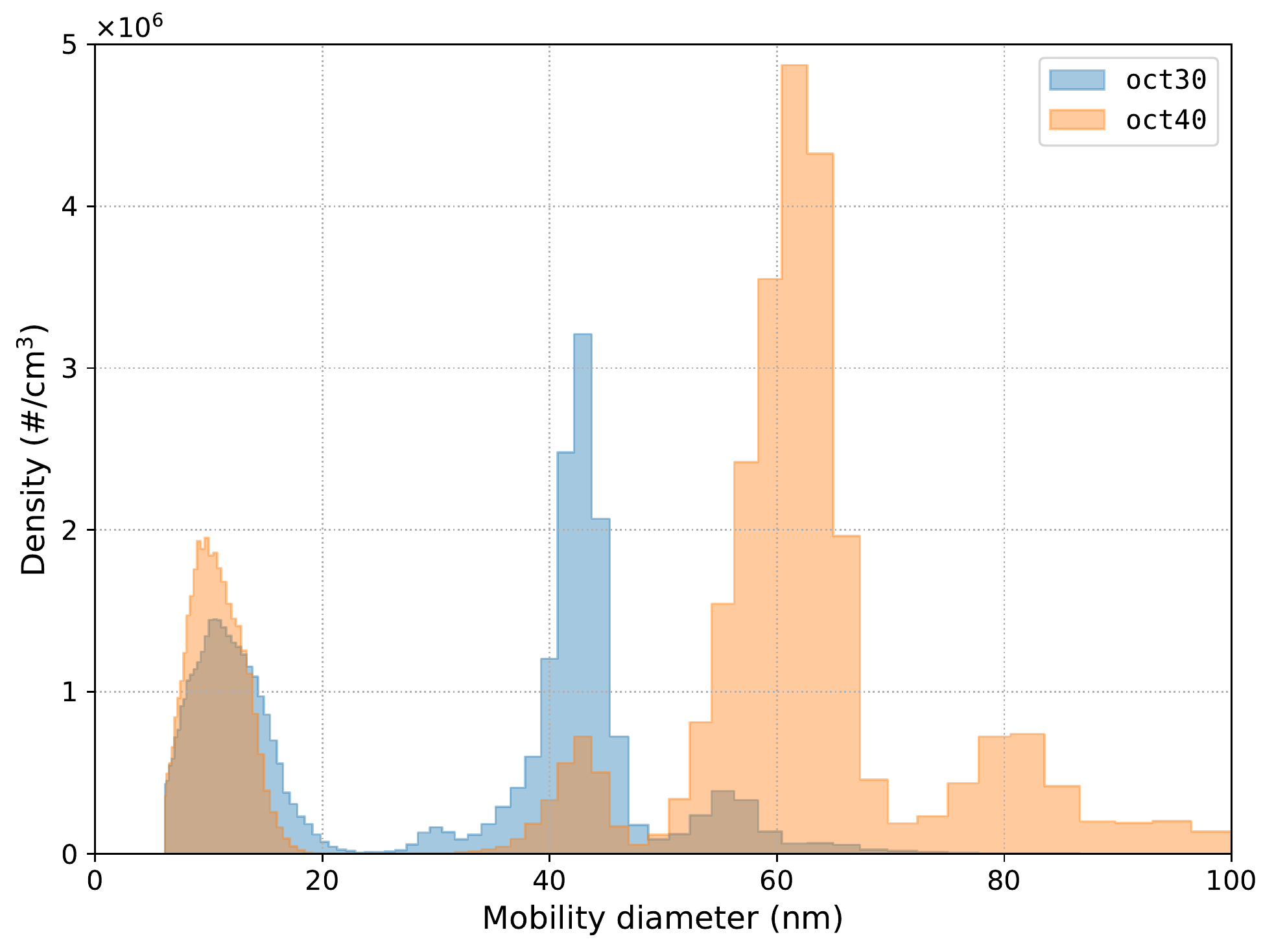} & 
    \includegraphics[width=0.45\textwidth]{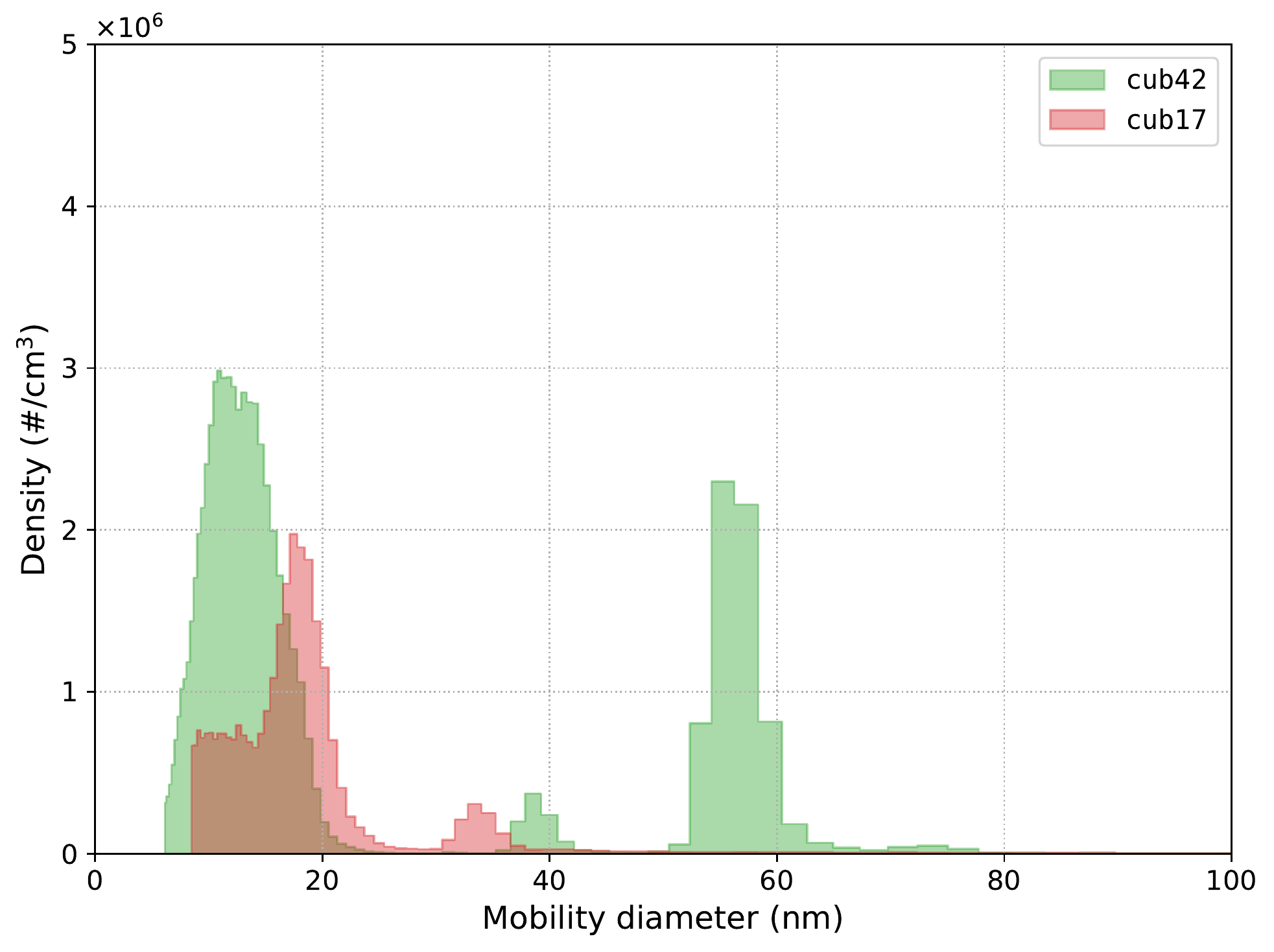} \\
    (a) & (b)
  \end{tabular}
  \caption{Differential mobility analysis. Representative size histograms for the (a) octahedral and (b) cubic samples obtained using the ES-DMA. The buffer peaks near \SI{10}{\nano\meter} are absent after passing through the aerodynamic lens to reach the interaction region and will, in any case, not be focused in the same region as the heavier AuNPs. Note that the ES-DMA measures the electrical mobility diameter which is dependent on both the size and shape of the particle~\cite{DeCarlo:2004}.}
  \label{fig:dma_sizing}
\end{figure}

\begin{figure}
  \centering
  \begin{tabular}{c}
    \includegraphics[width=\textwidth]{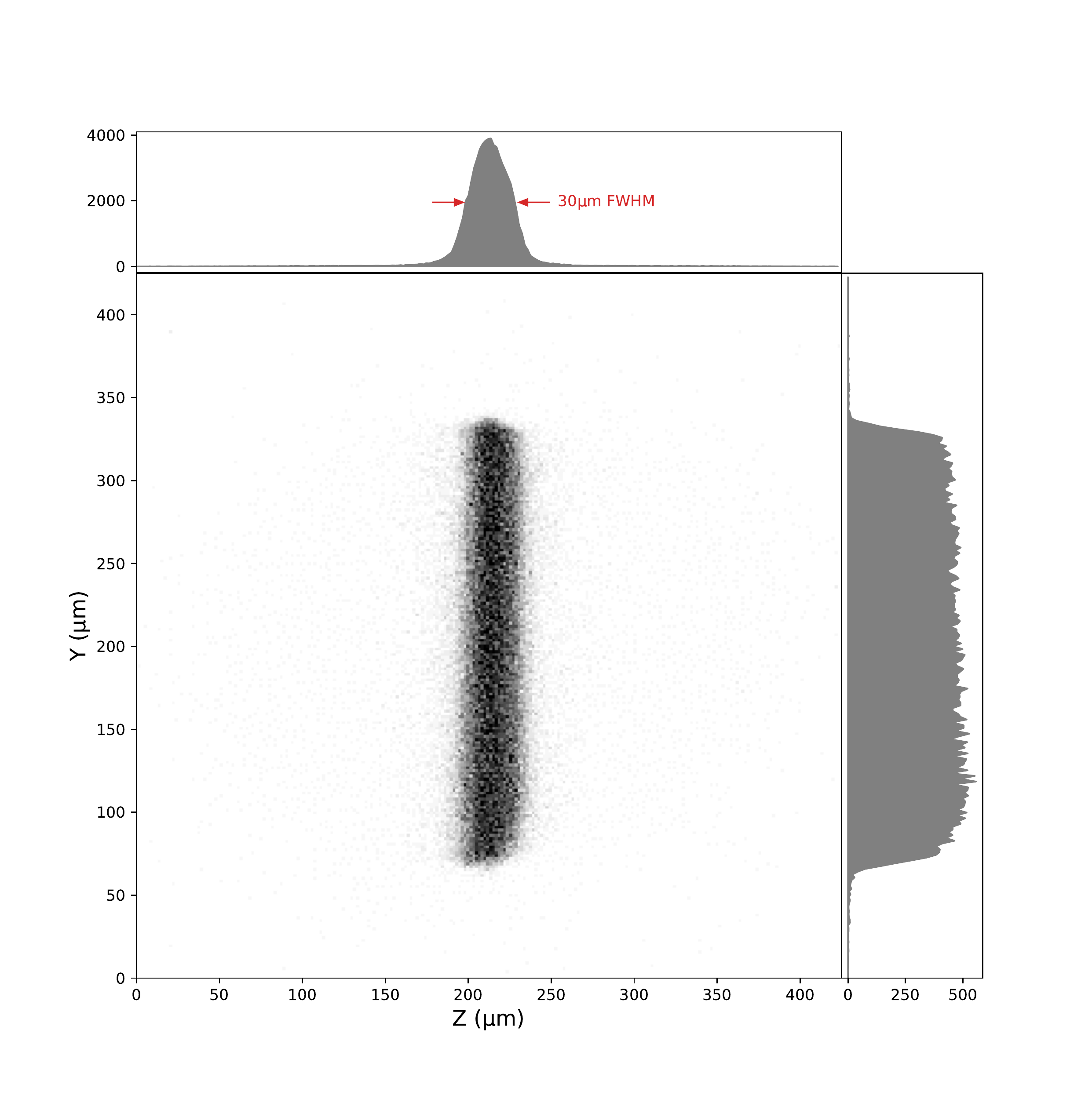} \\
    \includegraphics[width=0.8\textwidth]{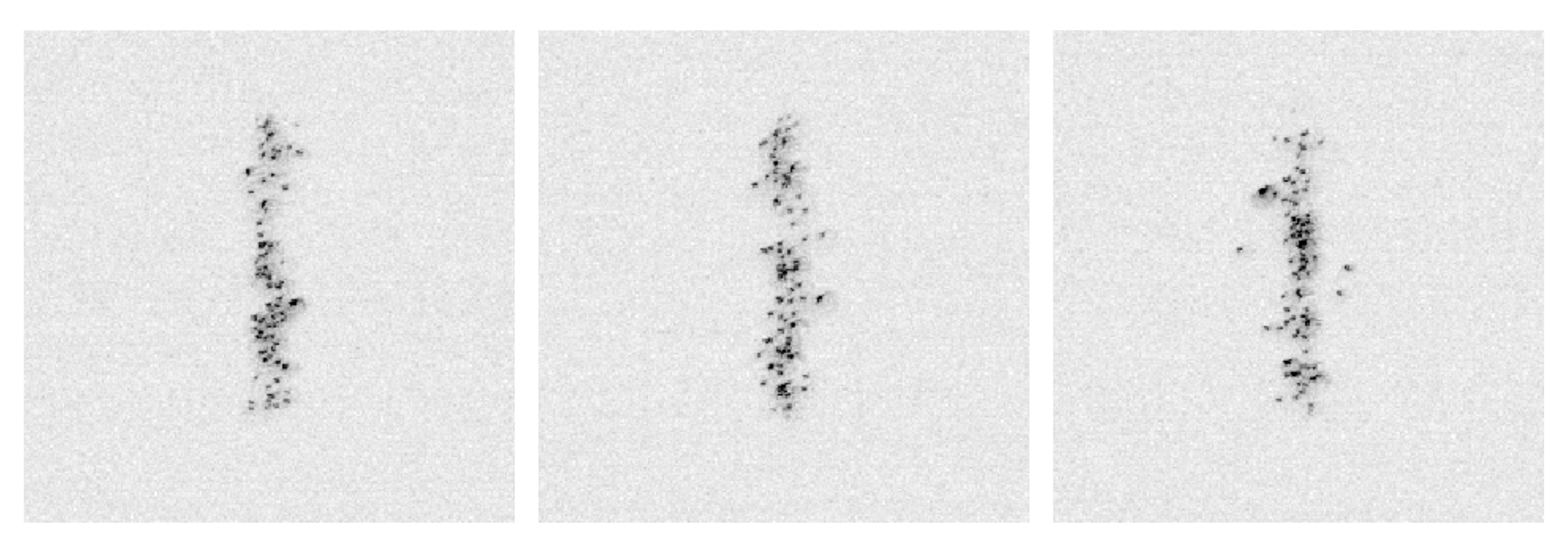}
  \end{tabular}
  \caption{Rayleigh scattering diagnostics. \emph{Top}: Histogram of detected particle positions in the interaction region for the \texttt{cub42} sample. The histogram was obtained from 1000 trains with one image per train. The X-ray beam propagates along the Z-axis and the Y-axis is vertical in the laboratory frame. The finite length of the particle beam in the vertical direction is due to the $\sim$\SI{250}{\micro\meter} optical laser spot size. The particle positions were detected using the \texttt{peak\_local\_max} function of Scikit Image~\cite{vanderwalt:2014} after appropriate background correction. \emph{Bottom row}: Three representative single background-corrected images from the same run.}
  \label{fig:rayleigh}
\end{figure}

\begin{figure}
  \centering
  \includegraphics[width=\textwidth]{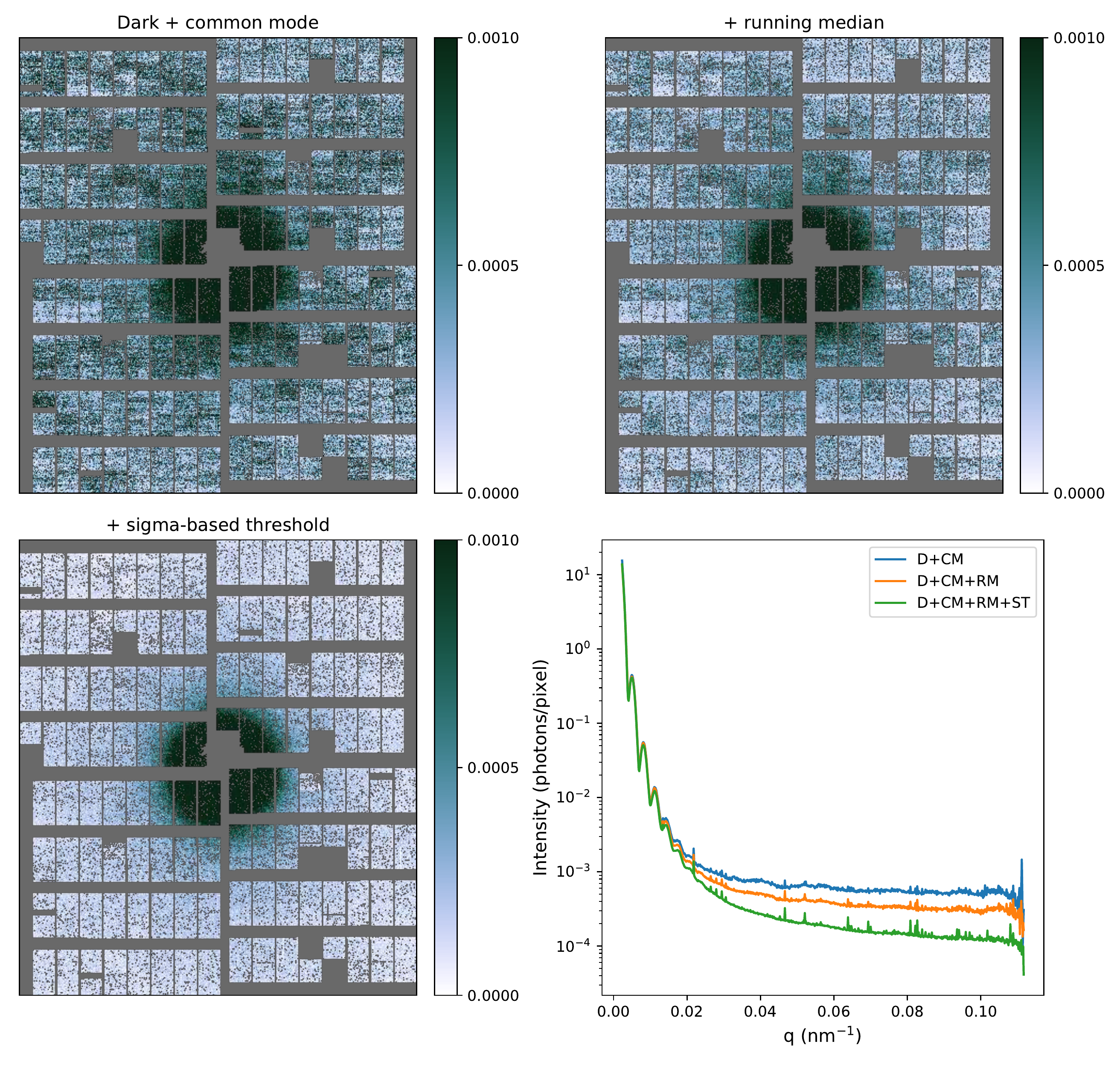}
  \caption{Detector corrections. Integrated photon-converted detector patterns for all \num{44800} hits in a single run with progressively more detector corrections. First only standard dark offset and common mode corrections were applied before thresholding at 0.7 of a photon. Next, the running median offset value over the last 128 trains was subtracted from each pixel before thresholding. Finally, a variable threshold was used for each memory cell depending upon the standard deviation (sigma) of the cell. All images have the same colour scale, saturating at $10^{-3}$ photons/pixel/frame. The dark gray background shows panel gaps and masked pixels. The fourth plot shows the radial average intensity for the three images.}
  \label{fig:detcorr}
\end{figure}

\clearpage

\subsection*{Supplementary Movie captions}
[Movies can be found here: https://owncloud.gwdg.de/index.php/s/ybXaOva83PUE4dQ]

\textbf{Supplementary Movie 1}: Intensity isosurfaces with varying levels for the the three larger samples.

\textbf{Supplementary Movie 2}: Rotating versions of electron density isosurfaces shown in Fig. 5(a).

\end{document}